\documentclass[%
reprint,
nofootinbib,
superscriptaddress,
amsmath,amssymb,
aps,
floatfix,
]{revtex4-2}

\usepackage[caption=false]{subfig}

\newcommand{\adsurl}[1]{\href{#1}{ADS}}

\newcommand{\jcap}{J.\ Cosmol.\ Astropart.\ Phys.}

\def\araa{ARA\&A}
\def\apj{ApJ}
 
\def\apjl{ApJL}                
\def\prd{Phys.~Rev.~D}
\def\physrep{Phys.~Rep.}
\def\aap{A\&A}

\def\prl{Phys.~Rev.~Lett.}
\def\na{New Astronomy}
\def\sovast{Soviet Astronomy}
\newcommand{\mnras}{Mon.\ Not.\ R.\ Astron.\ Soc.}

\newcommand\ba{\begin{eqnarray}}
    \newcommand\ea{\end{eqnarray}}
\newcommand\be{\begin{equation}}
    \newcommand\ee{\end{equation}}

\newcommand\gsim{ \lower .75ex \hbox{$\sim$} \llap{\raise .27ex \hbox{$>$}} }
\newcommand\lsim{ \lower .75ex \hbox{$\sim$} \llap{\raise .27ex \hbox{$<$}} }

\newcommand{\vx}{{\mathbf{x}}}

\newcommand{\vl}{{\mathbf{l}}}
\newcommand{\vL}{\mathbf{L}}

\usepackage{simpler-wick}
\usepackage{simplewick}

\usepackage{graphicx}
\usepackage{dcolumn}
\usepackage{bm}

\usepackage{hyperref}

\usepackage{xcolor}

\begin{document}
    
    
    \title{The impact of extragalactic foregrounds \\ on internal delensing  of CMB B-mode polarization}
    
    \author{Antón Baleato Lizancos}
    \email{a.baleatolizancos@berkeley.edu}
    \affiliation{Berkeley Center for Cosmological Physics, Department of Physics, University of California, Berkeley, CA 94720, USA}%
    \affiliation{Lawrence Berkeley National Laboratory, One Cyclotron Road, Berkeley, CA 94720, USA}
    \author{Simone Ferraro}%
    \email{sferraro@lbl.edu}
    \affiliation{Lawrence Berkeley National Laboratory, One Cyclotron Road, Berkeley, CA 94720, USA}
    \affiliation{Berkeley Center for Cosmological Physics, Department of Physics, University of California, Berkeley, CA 94720, USA}%
    
    \date{\today}
    
    \begin{abstract}
        The search for primordial $B$-mode polarization of the CMB is limited by the sample variance of $B$-modes produced at later times by gravitational lensing. Constraints can be improved by `delensing': using some proxy of the matter distribution to partially remove the lensing-induced $B$-modes. Current and soon-upcoming experiments will infer a matter map --- at least in part --- from the temperature anisotropies of the CMB. These reconstructions are contaminated by extragalactic foregrounds: radio-emitting galaxies, the cosmic infrared background, or the Sunyaev--Zel'dovich effects. Using the Websky simulations, we show that the foregrounds add spurious power to the angular auto-spectrum of delensed $B$-modes via non-Gaussian higher-point functions, biasing constraints on the tensor-to-scalar ratio, $r$. We consider an idealized experiment similar to the Simons Observatory, with no Galactic or atmospheric foregrounds. After removing point sources detectable at 143\,GHz and reconstructing lensing from CMB temperature modes $l<3500$ using a Hu-Okamoto quadratic estimator (QE), we infer a value of $r$ that is $1.5\,\sigma$ higher than the true $r=0$. Reconstructing instead from a minimum-variance ILC map only exacerbates the problem, bringing the bias above $3\,\sigma$. When the $TT$ estimator is co-added with other QEs or with external matter tracers, new couplings ensue which partially cancel the diluted bias from $TT$. We provide a simple and effective prescription to model these effects. In addition, we demonstrate that the point-source-hardened or shear-only QEs can not only mitigate the biases to acceptable levels, but also lead to lower power than the Hu-Okamoto QE after delensing. Thus, temperature-based reconstructions remain powerful tools in the quest to measure $r$.
    \end{abstract}
    
    \maketitle
    
    
    \section{\label{sec:intro} Introduction}
    The $B$-mode of polarization of the cosmic microwave background (CMB) is a powerful probe of the very early Universe. To leading order, primordial $B$-modes can only be produced by tensor fluctuations (gravitational waves) in the pre-recombination plasma, so --- unlike the temperature ($T$) or $E$-mode anisotropies --- they are not obscured by the cosmic variance of scalar fluctuations~\cite{Seljak:1996gy, Kamionkowski:1996zd}. $B$-modes are therefore a promising avenue to detect tensor modes, which are generically produced~\cite{Polnarev:1985} in theories of cosmic inflation~\cite{ref:guth_81, ref:linde_82, ref:albrecht_steinhardt_82, ref:starobinsky_79}, but not in many of its alternatives~\cite{ref:khoury_et_al_03}. A detection of primordial $B$-modes would reveal the amplitude of the power spectrum of tensor fluctuations, typically parametrized by the tensor-to-scalar ratio of primordial fluctuation power, currently constrained to be $r<0.036$~\cite{ref:BK_21} ($95\%$\,confidence level, at a pivot scale of $0.05\,\mathrm{Mpc}^{-1}$), and open the door to deeper investigations of inflation and fundamental Physics (see, e.g.~\cite{ref:s4_science_book}).
    \begin{figure}
        \centering
        \includegraphics[width=\columnwidth]{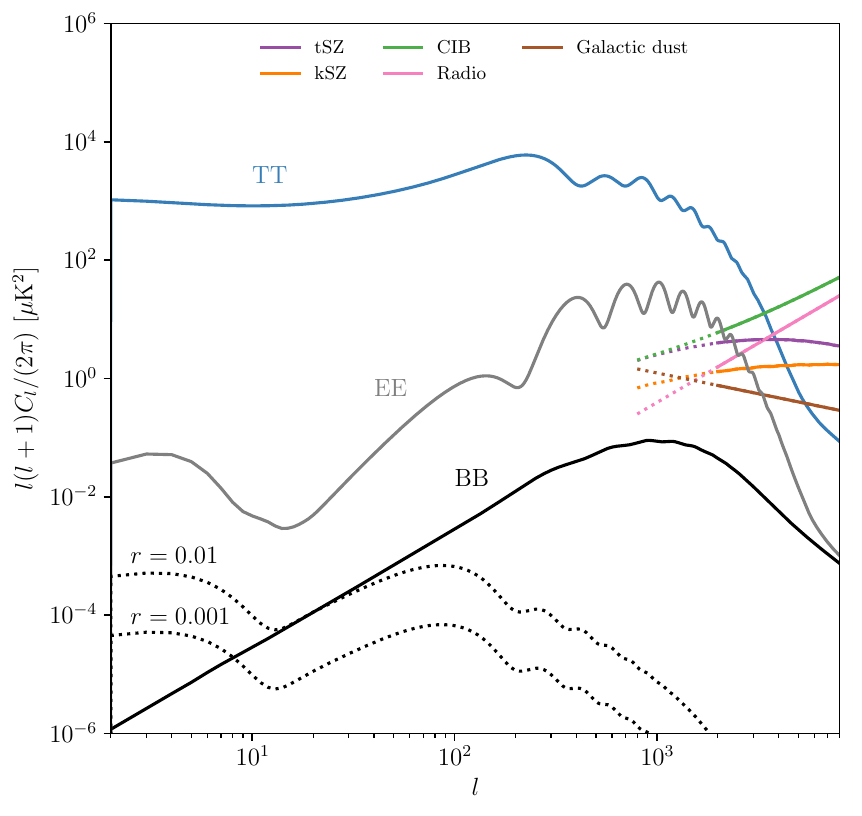}
        \caption{Angular power spectra of the lensed CMB temperature (blue), $E$-mode (gray) and $B$-mode polarization (black) in the $\Lambda$CDM cosmological model of~\cite{ref:planck_params_18}. We emphasize two points: i) the primordial $B$-modes (black, dotted) peak on degree scales, but they are obscured by the contribution from lensing (black, solid), and ii) on small scales, extragalactic foregrounds make sizeable contributions to the temperature fluctuations (to avoid clutter, we plot them only at large $l$). Note that although the primordial $B$-mode signal is enhanced at $l<10$ --- the so-called `reionization bump' --- this is exceedingly difficult to measure, especially from the ground~\cite{ref:s4_science_book}.}
        \label{fig:cls}
    \end{figure}

    Experimental constraints show the primordial $B$-mode signal is very small; the black, dotted curves in figure~\ref{fig:cls} are a theoretical calculation of its power spectrum for values of $r$ compatible with observations. Making a detection therefore entails overcoming various formidable challenges: from the development of high-sensitivity detectors and readout systems, to the accurate characterization of foreground emission and the tight control of experimental systematics (see, e.g.~\cite{ref:kamionkowski_kovetz_21}).
    
    In addition to all these `local' challenges, there is a major one originating beyond our Galaxy. As they travel towards us from the surface of last scattering, CMB photons see their path deflected by the gravitational pull of the matter distribution of the Universe --- they are gravitationally `lensed'; see~\cite{ref:lewis_challinor_review} for a review. This converts part of the $E$-mode polarization generated by scalars into $B$-modes, an effect first detected by~\cite{ref:hanson_13}. The angular power spectrum of these lensing $B$-modes, the black curve in figure~\ref{fig:cls}, resembles that of white noise with $\Delta_{\rm P}\approx5\, \mu \text{K\,arcmin}$ on the large angular scales where the primordial signal is expected to be strongest. Crucially, the variance associated with this lensing component hinders searches for a primordial $B$-mode signal produced by tensor fluctuations~\cite{ref:zaldarriaga_seljak_98}.
    
    The sample variance induced by lensing can be partially removed by estimating the specific realization of lensing $B$-modes present in the sky, and either removing them from map-level observations of the large-scale CMB polarization, or folding this information into a realization-dependent model around which to build a likelihood for $r$. This is known as `delensing' the $B$-modes~\cite{ref:knox_2002, Seljak:2003pn, ref:carron_19, ref:millea_et_al_20}. $B$-mode delensing has already been demonstrated on data~\cite{ref:carron_17, ref:manzotti_delensing, Planck2018:lensing, ref:polarbear_delensing_19, ref:han_20, ref:bicep_delensing}, and shown to improve constraints on $r$~\cite{ref:bicep_delensing}. In fact, lensing is thought to be the single largest source of uncertainty in the most recent BICEP/Keck analysis~\cite{ref:BK_21}. Going forward, experiments seeking to better constrain $r$ will rely on extensive delensing (e.g.,~\cite{ref:SO_delensing_paper, ref:bharat_forecasts, ref:diego-palazuelos_et_al_20, ref:S4_pgw_forecasts_22, ref:BK_moriond_22}).
    
\begin{figure}
        \centering
        \includegraphics[width=\columnwidth]{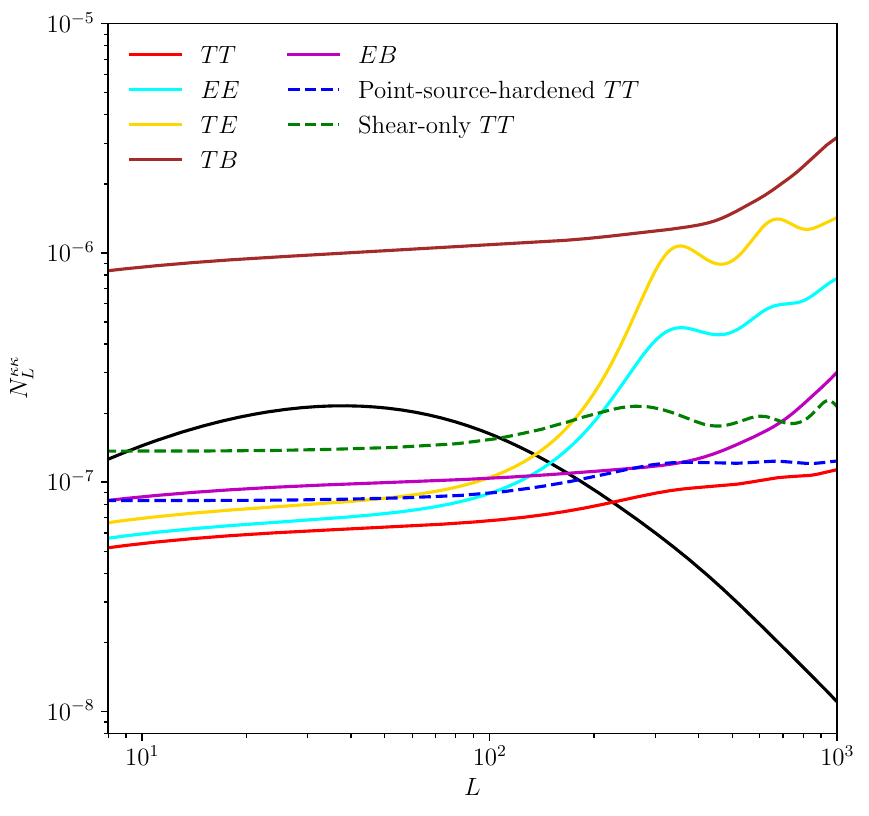}
        \caption{Theoretical reconstruction noise of various quadratic estimators of lensing, for the experiment described in section~\ref{sec:sims}, similar to the 143\,GHz channel of the large-aperture telescope of SO. These reconstructions use CMB anisotropies in the range $2<l<3000$, except for the $EB$ estimator, for which $300<l<3000$, as required to avoid delensing bias~\cite{ref:teng_11, ref:namikawa_17, ref:baleato_20_internal}. The noise is to be compared to the signal $C_L^{\kappa\kappa}$, shown in black. For SO, the $TT$ of Hu and Okamoto~\cite{ref:hu_okamoto_02} has the highest signal-to-noise of any quadratic estimator.}
        \label{fig:QE_noise}
    \end{figure}

    To estimate the realization of lensing $B$-modes in the sky, we need both high-resolution observations of the $E$-mode polarization and some proxy of the lensing potential~\cite{ref:smith_12_external, ref:limitations_paper}. The latter ingredient can be obtained internally, by reconstructing lensing maps from the CMB itself~\cite{ref:hu_okamoto_02, ref:hirata_seljak_pol, ref:carron_lewis_map_17, ref:millea_et_al_20}; externally, using tracers of the large-scale structure (which correlate with CMB lensing)~\cite{ref:smith_12_external, ref:sherwin_15, ref:yu_17}; or by combining both types of tracers (e.g.,~\cite{ref:SO_delensing_paper}). In this paper we focus on internal reconstructions, which are ultimately poised to return the highest-fidelity tracer of CMB lensing (see, e.g.,~\cite{ref:manzotti_delensing}), and will be an indispensable element of upcoming delensing analyses. In particular, we restrict our investigation to quadratic estimator (QE) reconstructions obtained from the CMB temperature anisotropies, since these offer the highest signal-to-noise lensing estimates out of all possible QEs for an experiment such as the upcoming Simons Observatory~\cite{ref:SO_science_paper} (see figure~\ref{fig:QE_noise}) and are likely to be used, albeit to a lesser extent, by other experiments such as the South Pole Observatory (see, e.g.,~\cite{ref:wu_et_al_19}) or CMB-S4~\cite{ref:s4_science_book}. Though QEs will ultimately be superseded by more sophisticated reconstruction algorithms which extract lensing information beyond leading order (e.g.,~\cite{ref:hirata_seljak_pol, ref:carron_lewis_map_17, ref:millea_et_al_20}), they are near-optimal for SO~\cite{ref:SO_science_paper}, and provide a transparent test bed for systematics that will be relevant to the more optimal --- and complex --- methods.

    The motivating fact for this work is that, as shown in figure~\ref{fig:cls}, a significant fraction of the temperature fluctuations in the small-scale microwave sky are produced by extragalactic `foreground' emission. These include the thermal Sunyaev-Zel'dovich effect~\cite{ref:sunyaev_zeldovich_72} (tSZ), a distortion of the CMB frequency spectrum in the direction of massive clusters where CMB photons inverse-Compton-scatter off hot electrons; the kinetic SZ~\cite{ref:sunyaev_zeldovich_80} (kSZ), a Doppler boost of those photons due to the bulk motion of electrons relative to the Hubble flow; the cosmic infrared background (CIB; see, e.g.~\cite{ref:hauser_dwek_01}), integrated emissions of interstellar dust in star-forming galaxies; and radio emission from active galactic nuclei (AGN), which appear as point sources in the CMB maps. All of these foregrounds are correlated with lensing (because they trace the matter distribution) and, importantly, obey non-Gaussian statistics. As a consequence, the auto- and cross-spectra of CMB lensing reconstructions can be biased at the level of several percent by the foregrounds' trispectra and their bispectra with CMB lensing~\cite{ref:amblard_et_al_04, ref:smith_07, ref:bleem_12, vanEngelen:2012va, ref:van_engelen_et_al_14, ref:osborne_et_al, ref:sailer_et_al, ref:madhavacheril_and_hill_18, ref:ferraro_hill_18, ref:schaan_ferraro_18, ref:sailer_et_al_21, ref:darwish_et_al_21, ref:han_sehgal_22}. Despite this being a well-known issue, the question of whether similar biases could affect B-mode delensing has remained unexplored to date.

    In this work, we show that not accounting for the non-Gaussian statistics of the extragalactic foregrounds can severely bias the delensing pipeline of an experiment similar to the Simons Observatory if the standard $TT$ quadratic estimator plays a part in the reconstructions. Interestingly, when the $TT$ estimator is coadded with other quadratic estimators and/or tracers of the matter distribution, new couplings arise which partially cancel the diluted bias from $TT$, leading to a further reduction in the overall bias, beyond the naive expectation based on the $TT$ weights alone. Fortunately, existing tools such as the point-source-hardened or shear-only quadratic estimators are relatively immune to foregrounds while retaining much of the delensing efficiency, and can be used to obtain overall-lower $B$-mode power than is possible with the standard HO QE. Moreover, we show that whatever spurious power there is after delensing can be accurately modeled using a simple prescription. 
    
    This paper is structured as follows. In section~\ref{sec:theory}, we lay out the theoretical framework behind $B$-mode delensing and CMB lensing reconstructions, and we use it to motivate our study of delensing biases from extragalactic foregrounds. In section~\ref{sec:methods}, we introduce the simulations and explain in detail the delensing pipeline used to assess these biases. The results of are presented in section~\ref{sec:results}, including a discussion of mitigation strategies and an extension to the case of multitracer delensing. We conclude with a discussion of the implications of this work in section~\ref{sec:conclusion}.

    \section{\label{sec:theory} Theoretical backbone}
    \subsection{Lensing and delensing of $B$-modes}
    We are interested in measuring the linear polarization of the CMB, the only type produced at leading order. Though this can in principle be captured by the $Q$ and $U$ Stokes parameters, polarization is a spin-2 field on the sphere, so $Q$ and $U$ are coordinate-dependent. It is therefore more convenient to work with the scalar $E$- and $B$-modes, which contain the same information and behave in unique ways under parity transformations: $E$-modes are even, while $B$-modes are odd. When the Stokes parameters are defined on a global $x$-$y$ basis, the two characterizations are related as    \footnote{There exist percent-level differences between results calculated using the flat- or spherical-sky mathematical formalisms; for example, in the power spectrum of lensing $B$-modes~\cite{ref:challinor_05}. Given the uncertainties present in other parts of our analysis --- particularly in the foreground models --- we deem it sufficient to work with the simpler flat-sky formalism. The results we obtain should still be qualitatively correct.}
    \begin{equation}
    (Q\pm i U)(\vx) = - \int \frac{d^2 \vl}{(2\pi)^2} \, \left[E(\vl)\pm i
  B(\vl)\right] e^{\pm 2i\psi_{\vl}} e^{i \vl\cdot\vx} \, ,
    \end{equation}
    where $\psi_{\vl}$ is the angle between $\vl$ and the $x$-direction.
    
    Lensing causes a remapping of the primary fields by a total deflection angle, $\bm{\alpha}$, as
    \begin{align}
        \tilde{T}(\bm{x}) &= T(\bm{x} + \bm{\alpha}(\vx))\,, \nonumber \\
        \tilde{Q}(\bm{x}) &= Q(\bm{x} + \bm{\alpha}(\vx))\, \quad \mathrm{and} \nonumber \\
        \tilde{U}(\bm{x}) &= U(\bm{x} + \bm{\alpha}(\vx))\,, \nonumber \\
    \end{align}
    where tildes denote lensed fields. Since the curl-like component of the deflection angle is orders of magnitude smaller than the gradient term~\cite{ref:hirata_seljak_pol, ref:pratten_and_lewis}, we may work simply with the lensing convergence, defined as $\kappa = -\frac{1}{2} \bm{\nabla} \cdot \bm{\alpha}$, and related to the lensing potential by $\kappa = -  \nabla^2 \phi/2$.

    To leading order, the $B$-mode polarization induced by gravitational lensing of $E$-modes can be written as
    \begin{align}\label{eqn:lensB}
        \tilde{B}(\bm{l}) &\approx \int \frac{d^2\bm{l}'}{(2\pi)^2} W(\bm{l},\bm{l}') E(\bm{l}')\kappa(\bm{l}-\bm{l}') \, ,
    \end{align}
    where
    \begin{equation}
        W(\bm{l},\bm{l}') = \frac{2 \bm{l}' \cdot (\bm{l}-\bm{l}')}{|\bm{l}-\bm{l}'|^2} \sin 2(\psi_{\bm{l}} - \psi_{\bm{l}'}) \, .
    \end{equation}
    On large angular scales, where the primordial $B$-mode power is expected to peak, the power spectrum of equation~\eqref{eqn:lensB} is an excellent approximation to the true, non-perturbative result~\cite{ref:challinor_05}. The reasons for this are subtle (see~\cite{ref:limitations_paper}) but they have a clear (and perhaps unintuitive) implication for delensing: a faithful lensing $B$-mode template can be constructed as
    \begin{align}\label{eqn:template}
        \hat{B}^{\mathrm{lens}}(\bm{l}) &= \int \frac{d^2\bm{l}'}{(2\pi)^2} W(\bm{l},\bm{l}') \mathcal{W}^{E}_{l'} \mathcal{W}^{\kappa}_{|\bm{l}-\bm{l}'|} \nonumber \\
        & \hphantom{= \int \frac{d^2\bm{l}'}{(2\pi)^2} f(\bm{l}, \bm{l}')} \times  E^{\mathrm{obs}}(\bm{l}')\hat{\kappa}(\bm{l}-\bm{l}')  \nonumber \\
        & \equiv g_l \left[ E^{\mathrm{obs}} \hat{\kappa}\right]\,,
    \end{align}
    as long as lensed --- not delensed or unlensed --- $E$-modes are used~\cite{ref:limitations_paper}.
    Here, $E^{\mathrm{obs}}$ are the observed $E$-modes, $\hat{\kappa}$ is an estimate of the convergence, and $\mathcal{W}^{E}$ and $\mathcal{W}^{\kappa}$ are Wiener filters for $E^{\mathrm{obs}}$ and $\hat{\kappa}$, respectively \footnote{As long as the fiducial spectra used in the Wiener filters are close to the truth, this deviation will only mildly reduce the delensing efficiency, for corrections to the power spectrum of delensed $B$-modes are second order in the error in the weights~\cite{ref:sherwin_15}.}. The expression above gives a lensing $B$-modes template that is effectively optimal until the fidelity of mass tracer and $E$-mode observations allow for delensing residuals of $O(1\%)$ of the original lensing $B$-mode power --- that is, beyond the era of CMB-S4~\cite{ref:limitations_paper}.
    
    Large-scale $B$-mode polarization can then be delensed by subtracting this template off from observations\footnote{It is common practice to delens by including the lensing template as an additional `channel' in a multi-frequency, cross-spectral pipeline (e.g.,~\cite{ref:bicep_delensing}); this approach can be shown to be equivalent to a map-level subtraction.},
    \begin{align}\label{eqn:Bdel}
        \hat{B}^{\mathrm{del}}(\bm{l}) = \tilde{B}^{\mathrm{obs}}(\bm{l}) -  \hat{B}^{\mathrm{lens}}(\bm{l})\,;
    \end{align}
    the power spectrum of $B$-modes after delensing is then
    \begin{equation}
    C_l^{BB, \mathrm{del}} = C_l^{BB} + N_l^{BB} + C_l^{BB, \mathrm{res}} +  C_l^{BB, \mathrm{fg}}\,,
    \end{equation}
    where $C_l^{BB}$, $N_l^{BB}$, $C_l^{BB, \mathrm{res}}$ and $C_l^{BB, \mathrm{fg}}$ are the angular power spectra of primordial/unlensed, experiment noise, residual lensing and foreground $B$-modes, respectively.
    
     The last of these components, the contribution from polarized foregrounds, is known to be a significant hurdle to searches for primordial $B$-modes and can come into play in a variety of ways. A well known challenge lies in the fact that, due to synchrotron radiation and thermal dust emission, our Galaxy is the source of bright, polarized light at CMB frequencies. At their minimum frequency of 70–90 GHz and on the angular scales relevant to $r$-science, these foregrounds are at a level comparable to a primordial signal with $r = 0.01$ – 0.1, depending on sky region (see, e.g.,~\cite{ref:krachmalnicoff_et_al_16} and references therein). In fact, current, high-sensitivity experiments already rely on foreground cleaning to achieve their target constraints (e.g.,~\cite{ref:planck_bicep_15}). In parallel, polarized, extragalactic foregrounds also need to be taken into account, particularly when targeting a recombination signal with $r < 0.01$ and using small- or medium-aperture telescopes with relatively-high confusion limits for point-source masking~\cite{ref:lagache_et_al_20}. Finally, it is also possible for foreground residuals present in the large-scale $B$-mode maps to couple with residuals in the matter tracer  and $E$-mode maps used to build the lensing $B$-mode template, biasing the delensing pipeline (e.g.,~\cite{ref:beck_et_al_18, ref:cib_delensing_biases}). All of these challenges are topics of active research, and we shall not consider them further in this paper. Instead, we will focus on how extragalactic foregrounds can bias lensing maps obtained from the CMB temperature anisotropies and propagate to biased inferences of $r$.

     Before we motivate our reasons for concern in section~\ref{sec:bias_explanation_singletr}, let us write out explicitly the terms that make up the angular power spectrum of residual lensing $B$-modes after delensing:
    \begin{equation}
        C_l^{BB, \mathrm{res}} = \tilde{C}_l^{BB} - 2\,C_l^{\tilde{B} \times \hat{B}^{\mathrm{lens}}} + C_l^{\hat{B}^{\mathrm{lens}} \times \hat{B}^{\mathrm{lens}}}\,,
    \end{equation}
    where $\tilde{C}^{BB}$ is the power spectrum of lensing $B$-modes, and we have defined the cross-spectrum between the true lensing $B$-modes and the template,
    \begin{align}
    C_l^{\tilde{B} \times \hat{B}^{\mathrm{lens}}} & = \langle \tilde{B}(\bm{l}) \hat{B}^{\mathrm{lens}}(\bm{l}') \rangle' \nonumber \\
    & = g_l\left[ \langle \tilde{B} E^{\mathrm{obs}} \hat{\kappa} \rangle \right]\, ,
    \end{align}
    with the prime following the angle bracket denoting that the delta function $\delta(\bm{l}-\bm{l}')$ has been removed. We have also written the auto-spectrum of the template as
    \begin{align}
        C_l^{\hat{B}^{\mathrm{lens}} \times \hat{B}^{\mathrm{lens}}} & = \langle \hat{B}^{\mathrm{lens}}(\bm{l}) \hat{B}^{\mathrm{lens}}(\bm{l}') \rangle' \nonumber \\
        &= \int \frac{d^2\bm{l}_1}{(2\pi)^2} W(\bm{l},\bm{l}_1) \mathcal{W}^{E}_{l_1} \mathcal{W}^{\kappa}_{|\bm{l}-\bm{l}_1|} \nonumber \\
        &\quad\times \int \frac{d^2\bm{l}_2}{(2\pi)^2} W(\bm{l}',\bm{l}_2) \mathcal{W}^{E}_{l_2} \mathcal{W}^{\kappa}_{|\bm{l}'-\bm{l}_2|} \nonumber\\
        & \qquad \times \langle E^{\mathrm{obs}}(\bm{l}_1)  \hat{\kappa} (\bm{l} - \bm{l}_1) E^{\mathrm{obs}}(\bm{l}_2) \hat{\kappa}(\bm{l} - \bm{l}_2)\rangle \nonumber \\
        & \equiv h_l\left[\langle E^{\mathrm{obs}}  \hat{\kappa} E^{\mathrm{obs}} \hat{\kappa}\rangle \right] \,.
    \end{align}
    In an ideal scenario, absent biases from foregrounds and otherwise, the residual lensing power is well modeled as~\cite{ref:smith_12_external, ref:sherwin_15, ref:baleato_20_internal}
    \begin{equation}\label{eqn:clbbdel_theory}
        C_l^{BB,\mathrm{res}} = \int \frac{d^2\bm{l}'}{(2\pi)^2} W^2(\bm{l},\bm{l}') C_{l'}^{EE}C_{|\bm{l}-\bm{l}'|}^{\kappa\kappa} \left[1- \mathcal{W}^{E}_{l'}\rho^2_{|\bm{l}-\bm{l}'|} \right] \, ,
    \end{equation}
    where $\mathcal{W}^{E}_{l}$ is a Wiener-filter for $E$-modes (assumed here and throughout to be diagonal, for simplicity) and 
    \begin{equation}\label{eqn:def_rho}
        \rho_L \equiv \frac{C_L^{\kappa \hat{\kappa}}}{\sqrt{C_L^{\kappa\kappa}C_L^{\hat{\kappa}\hat{\kappa}}}}\,
    \end{equation}
     is the correlation coefficient between the mass tracer and CMB lensing.

    \subsection{Quadratic estimators of lensing}\label{sec:qes}
    In order to build the lensing $B$-mode template of equation~\eqref{eqn:template}, we need an estimate of the convergence map responsible for the deflections. In this paper, we focus on quadratic estimators derived from CMB temperature fields, as these will play an important role in current and upcoming experiments and are more prone to contamination from extragalactic foregrounds than estimators relying on polarization~\cite{ref:smith_cmbpol}. A general $TT$ quadratic estimator of lensing can be written as
    \begin{align}\label{eqn:qe}
        \hat{\kappa}(\bm{L}) &= N^{\kappa\kappa}(\bm{L})\int \frac{d^2\bm{l}}{(2\pi)^2} F^{\mathrm{\kappa}}(\bm{l}, \bm{L} -\bm{l}) T^{\mathrm{obs}}(\bm{l})  T^{\mathrm{obs}}(\bm{L} -\bm{l})\,.
    \end{align}
    If the normalization is chosen to be
    \begin{align}\label{eqn:qe_noise}
        \left[N^{\kappa\kappa}(\bm{L})\right]^{-1} &= \int \frac{d^2\bm{l}}{(2\pi)^2} F^{\mathrm{\kappa}}(\bm{l}, \bm{L} -\bm{l}) f^{\kappa}(\bm{l},\bm{L} -\bm{l})\,,
    \end{align}
    then to leading order the estimator has unit response to lensing, and $N^{\kappa\kappa}$ gives the Gaussian component of the reconstruction noise\footnote{In the idealized limit of homogeneous and isotropic noise}; this is the expression we used to calculate the various noise curves in figure~\ref{fig:QE_noise}. In the expression above,
    \begin{align}\label{eqn:fkappa}
        f^{\kappa}(\bm{l},\bm{L} -\bm{l}) = \frac{2 \bm{L}}{L^2} \cdot \left[ \bm{l} \tilde{C}_{l}^{TT} + (\bm{L} - \bm{l}) \tilde{C}_{|\bm{L}-\bm{l}|}^{TT}\right]\,,
    \end{align}
    where $\tilde{C}_{l}^{TT}$ is the lensed CMB temperature angular power spectrum\footnote{Ref.~\cite{ref:lewis_11} showed that $\tilde{C}_{l}^{TT}$ is a better approximation to the non-perturbative weights than $C_{l}^{TT}$.}. On the other hand, one is free to choose the form of $F^{\mathrm{\kappa}}(\bm{l}, \bm{L} -\bm{l})$ to satisfy some desired property. The standard Hu-Okamoto (HO) quadratic estimator of Ref.~\cite{ref:hu_okamoto_02} has
    \begin{align}\label{eqn:F_HO}
        F^{\mathrm{\kappa}, \mathrm{HO}}(\bm{l}, \bm{L} -\bm{l}) =  \frac{f^{\kappa}(\bm{l},\bm{L} -\bm{l})}{2C^{\mathrm{tot}}_{l}C^{\mathrm{tot}}_{|\bm{L}-\bm{l}|}}\,,
    \end{align}
    which results in the minimum variance estimator of lensing that is quadratic in the temperature fields.

    We will be comparing this so-called `standard' QE to alternative quadratic estimators which are by design more robust to biases from extragalactic foregrounds. One of the alternative estimators we consider is the shear-only estimator of Ref.~\cite{ref:schaan_ferraro_18}, defined by the weights
    \begin{align}\label{eqn:F_shear}
        F^{\mathrm{\kappa}, \mathrm{shear}}(\bm{l}, \bm{L} -\bm{l}) = \frac{\cos \left[2(\psi_{\vL}-\psi_{\vl})\right]\tilde{C}_l^{TT}}{2C^{\mathrm{tot}}_{l}C^{\mathrm{tot}}_{|\bm{L}-\bm{l}|}} \frac{d \ln \tilde{C}_l^{TT}}{d \ln l}\,.
    \end{align}
    The reason for this nomenclature is that in the limit where large-scale lenses are reconstructed from much smaller anisotropies the estimator can be shown to extract information only from the shear. This is a desirable property that bestows upon the estimator a high degree of immunity to extragalactic foregrounds, which are approximately azimuthally-symmetric --- their imprint on the CMB is therefore degenerate with the lensing magnification, but not the shear.
     
     We also consider a bias-hardened estimator~\cite{ref:osborne_et_al, ref:sailer_et_al}, $\hat{\kappa}^\mathrm{BH}_{\bm{L}}$, which by design has zero response to point sources at leading order; it is defined as
    \begin{equation}
    \begin{pmatrix}
     \hat{\kappa}^\mathrm{BH}_{\bm{L}}\\
     \hat{s}^{\mathrm{BH}}_{\bm{L}}
    \end{pmatrix} =
    \begin{pmatrix}
    1 & N^\kappa_{\bm{L}}\mathcal{R}_{\bm{L}}\\
    N^{s}_{\bm{L}}\mathcal{R}_{\bm{L}} & 1
    \end{pmatrix}^{-1}
    \begin{pmatrix}
     \hat{\kappa}_{\bm{L}}\\
     \hat{s}_{\bm{L}}
    \end{pmatrix}
    ,
    \label{eq:bias-hardened_k_s}
    \end{equation}
    where 
    \begin{align}
        \mathcal{R}_{\bm{L}} &= \int \frac{d^2\bm{l}}{(2\pi)^2} \frac{f^{\kappa}(\bm{l}, \bm{L} -\bm{l}) f^{s}(\bm{l},\bm{L} -\bm{l})}{2C^{\mathrm{tot}}_l C^{\mathrm{tot}}_{|\bm{L}-\bm{l}|}  }\,,
    \end{align}
    and $\hat{s}_{\bm{L}}$ and $N^{s}_{\bm{L}}$ are defined by analogy with $\hat{\kappa}_{\bm{L}}$ and $N^{\kappa}_{\bm{L}}$ --- they can be obtained from equations~\eqref{eqn:qe}, \eqref{eqn:qe_noise} and \eqref{eqn:F_HO} by replacing $\kappa$ with $s$ in the superscripts. For point sources, $f^{s}(\bm{l},\bm{L} -\bm{l})$=1.

    \subsection{Delensing bias from foreground non-Gaussianity} \label{sec:bias_explanation_singletr}
    The foregrounds are non-Gaussian and correlated with the matter distribution that gravitationally lenses the CMB. This causes internal reconstructions of CMB lensing to be biased, in turn biasing the power spectrum of delensed $B$-modes. To isolate this effect, we define the bias as the difference in delensed-$B$-mode power between two pipelines which only differ by the statistical properties of the foregrounds that enter the QE. In this way, we can ensure that the leading-order reconstruction noise is the same in the two cases\footnote{The Gaussian reconstruction noise is the same if the foregrounds have matching power spectra, but the non-Gaussian noise terms --- $N^{(1)}$ and higher-order --- can in principle differ. We ignore this subtlety, since the Gaussian term dominates across the scales where the reconstruction is signal-dominated (e.g.,~\cite{ref:hanson_et_al_11}).}, and thus that the delensing efficiency is only changed by the decorrelation induced by the foreground non-Gaussianity. Mathematically, this definition corresponds to
    \begin{align}\label{eqn:bias}
        \Delta C_l^{BB, \mathrm{res}} \equiv \langle |\tilde{B} - g_l\left[E^{\mathrm{obs}} \hat{\kappa}\left[f_{X} +s_{X}^{\mathrm{NG}}, f_{Y} +s_{Y}^{\mathrm{NG}}\right]\right]|^{2} \rangle \nonumber \\
        - \langle |\tilde{B} - g_l\left[E^{\mathrm{obs}} \hat{\kappa}\left[f_{X} +s_{X}^{\mathrm{G}}, f_{Y} +s_{Y}^{\mathrm{G}}\right]\right]|^{2} \rangle \,.
    \end{align}
    where $s_{X}$ is the extragalactic foreground contribution to leg $X$ of the $XY$ QE, with the (NG) G superscript denoting that the foreground map in question is (non-) Gaussian. On the other hand, $f_{X}$ refers to all the other non-foreground contributions to leg $X$: lensed CMB, experiment noise, etc.

    In this work, we assume that extragalactic foregrounds are negligible in polarization, and set $s\equiv s_{T}$ while $s_{E}=s_{B}=0$. This is a very good approximation given current sensitivity requirements because extragalactic foregrounds are expected to be polarized only at the percent level, if at all. In the future, it might be necessary to study the additional couplings that arise when the $B$- and $E$-modes in the expression above receive contributions from extragalactic foregrounds, and to investigate also the case of polarization-based reconstructions, which we do not consider here. However, the present treatment should be highly accurate for SO, particularly with the extension to the case of multitracer delensing that we provide in section~\ref{sec:multitracer}.
    
    The total bias described by equation~\eqref{eqn:bias} can be split into two parts:
    \begin{equation}\label{eqn:bias_TT}
        \Delta C_l^{BB, \mathrm{res}} = -2\,\Delta C_l^{\tilde{B} \times \hat{B}^{\mathrm{lens}}} + \Delta C_l^{\hat{B}^{\mathrm{lens}}\times \hat{B}^{\mathrm{lens}}} \,,
    \end{equation}
    the bias to the cross-spectrum of template and lensing $B$-modes, and the bias to the template auto-spectrum. When delensing with a $TT$ estimator, these can include contributions such as\footnote{For notational economy, we drop from here on out the subscripts labeling the inputs to each QE leg. Though the new notation does not suggest this explicitly, the reader should remember that $s$ is a placeholder for the combination of all extragalactic foregrounds --- tSZ, kSZ, radio or CIB --- once the statistical properties (Gaussian or non-Gaussian) have been specified. This is an important point to encompass mixed biases sourced, for example, via the CIB-tSZ correlation.}
    \begin{equation}\label{eqn:bias_couplings_cross}
    \Delta C_l^{\tilde{B} \times \hat{B}^{\mathrm{lens}}} \supset g_l \left[ \langle \tilde{B} \tilde{E} \hat{\kappa}^{TT} \left[s^{\mathrm{NG}},s^{\mathrm{NG}}\right] \rangle_{\mathrm{c}} \right] \,,
    \end{equation}
    and
    \begin{align}\label{eqn:bias_couplings_auto}
        \Delta & C_l^{\hat{B}^{\mathrm{lens}}\times  \hat{B}^{\mathrm{lens}}} \supset  \nonumber \\
        & \supset \wick[offset=1.5em]{ 2\,h_l \left[ \langle  \c {E}^{\mathrm{obs}} \hat{\kappa}^{\mathrm{TT}} \left[\tilde{T},\tilde{T}\right]  \c {E}^{\mathrm{obs}} \hat{\kappa}^{\mathrm{TT}} \left[s^{\mathrm{NG}},s^{\mathrm{NG}}\right]\rangle_{\mathrm{c}}\right]} \nonumber \\
        & \wick[offset=1.5em]{ + 4\,h_l \left[ \langle \c {E}^{\mathrm{obs}} \hat{\kappa}^{\mathrm{TT}} \left[\tilde{T},s^{\mathrm{NG}}\right] \c {E}^{\mathrm{obs}} \hat{\kappa}^{\mathrm{TT}} \left[\tilde{T},s^{\mathrm{NG}}\right]\rangle_{\mathrm{c}}\right]} \nonumber \\
        & \wick[offset=1.5em]{ + h_l \left[ \langle \c {E}^{\mathrm{obs}} \hat{\kappa}^{\mathrm{TT}} \left[s^{\mathrm{NG}},s^{\mathrm{NG}}\right] \c {E}^{\mathrm{obs}} \hat{\kappa}^{\mathrm{TT}} \left[s^{\mathrm{NG}},s^{\mathrm{NG}}\right]\rangle_{\mathrm{c}}\right]} \,.
    \end{align}
    %
    In our notation, a Gaussian contraction of the fields connected by over-bars is to be taken first, and this is then to be multiplied by the connected $n$-point function of the remaining fields inside the angle brackets. Note that all of these terms vanish if the foregrounds are Gaussian. We only show a subset of all possible contributions to $\Delta C_l^{\hat{B}^{\mathrm{lens}}\times  \hat{B}^{\mathrm{lens}}}$, those that we believe should dominate based on the coupling structure of the weights inside the integrand\footnote{We follow the arguments in appendix~A of~\cite{ref:baleato_20_internal}. First, we rank terms according to their order in lensing; since the foregrounds are highly correlated with lensing, we take $\kappa$ and $s$ to be equivalent in this counting exercise. Then, we consider how different couplings affect the volume of multipole space over which the integrals are allowed to accumulate their signal. More tightly coupled integrands are likely to produce smaller results upon integration, so we rank them less highly in our priority list.}.  It is couplings structurally identical to these we have retained (without the foregrounds) that dominate in the nominal calculation of the residual lensing $B$-mode spectrum and simplify to equation~\eqref{eqn:clbbdel_theory} ~\cite{ref:baleato_20_internal, ref:namikawa_17}. A complete taxonomy of possible contributions is provided in appendix~\ref{appendix:TT_biases}.
    
    The couplings we have retained take a particularly simple analytic form. To leading order, they give
    \begin{align}\label{eqn:full_theory_cross}
        \Delta C_l^{\tilde{B}\times \hat{B}^{\mathrm{lens}}} = \int \frac{d^2\bm{l}'}{(2\pi)^2} W^2(\bm{l},\bm{l}') & \mathcal{W}^{E}_{l'} \mathcal{W}^{\kappa}_{|\bm{l}-\bm{l}'|} \nonumber \\
        & \times C^{EE}_{l'} \Delta C^{\kappa \hat{\kappa}}_{\bm{l}-\bm{l}'}\,,
    \end{align}
    and
    \begin{align}\label{eqn:full_theory_auto}
        \Delta C_l^{\hat{B}^{\mathrm{lens}}\times \hat{B}^{\mathrm{lens}}} = \int \frac{d^2\bm{l}'}{(2\pi)^2} W^2(&\bm{l},\bm{l}') \left(\mathcal{W}^{E}_{l'} \mathcal{W}^{\kappa}_{|\bm{l}-\bm{l}'|} \right)^2 \nonumber \\
        & \times C^{EE, \mathrm{tot}}_{l'}
        \Delta C^{\hat{\kappa} \hat{\kappa}}_{\bm{l}-\bm{l}'}\,,
    \end{align}
    where $\Delta C^{\hat{\kappa} \hat{\kappa}}$ and $\Delta C^{\kappa \hat{\kappa}}$ are, respectively, the foreground-induced biases to the reconstruction's auto- and cross-spectra with the true CMB lensing convergence.

    We are now in a position to make an explicit connection between the couplings in equations~\eqref{eqn:bias_couplings_cross} and~\eqref{eqn:bias_couplings_auto} and the biases studied extensively in the context of the auto- and cross-spectra of CMB lensing reconstructions --- e.g.~\cite{ref:amblard_et_al_04, ref:osborne_et_al, ref:van_engelen_15, ref:ferraro_hill_18, ref:schaan_ferraro_18, ref:sailer_et_al, ref:sailer_et_al_21, ref:darwish_et_al_21}. To leading order in lensing, the trispectrum in equation~\eqref{eqn:bias_couplings_cross} reduces to a product of the unlensed $E$-mode spectrum times 
    \begin{equation}\label{eqn:bispec_bias}
        \langle \kappa \,\hat{\kappa}^{TT} \left[s^{\mathrm{NG}},s^{\mathrm{NG}}\right] \rangle_{\mathrm{c}}\,.
    \end{equation}
    This object, essentially a $\langle \kappa s s\rangle$ bispectrum, is known to bias cross-correlations of CMB lensing reconstructions with any tracer of the large-scale structure. In other words, it sources $\Delta C^{\kappa \hat{\kappa}}$ in equation~\ref{eqn:full_theory_cross}. 

    On the other hand, the terms in equation~\eqref{eqn:bias_couplings_auto} feature a Gaussian contraction of the $E$-mode legs across templates, together with a connected four-point function of the fields involved in the lensing reconstruction. The latter produces a $\Delta C^{\hat{\kappa} \hat{\kappa}}$ in equation~\eqref{eqn:full_theory_auto}. Heuristically, the first line contains
    \begin{equation}\label{eqn:prim_bispec}
        \langle \hat{\kappa}^{\mathrm{TT}} \left[\tilde{T},\tilde{T}\right] \hat{\kappa}^{\mathrm{TT}} \left[s^{\mathrm{NG}},s^{\mathrm{NG}}\right]\rangle_{\mathrm{c}} \,,
    \end{equation}
    so it is related to the `primary bispectrum bias'\footnote{To leading order in lensing, terms~\eqref{eqn:prim_bispec} and~\eqref{eqn:sec_bispec} both involve $\langle \kappa s s\rangle$ bispectra, hence the nomenclature.} discussed in the literature; the second line is a function of
    \begin{equation}\label{eqn:sec_bispec}
        \langle \hat{\kappa}^{\mathrm{TT}} \left[\tilde{T},s^{\mathrm{NG}}\right] \hat{\kappa}^{\mathrm{TT}} \left[\tilde{T},s^{\mathrm{NG}}\right]\rangle_{\mathrm{c}}\,,
    \end{equation}
    so it can be associated with the `secondary bispectrum bias'; and the third line is sourced by
    \begin{equation}
        \langle \hat{\kappa}^{\mathrm{TT}} \left[s^{\mathrm{NG}},s^{\mathrm{NG}}\right] \hat{\kappa}^{\mathrm{TT}} \left[s^{\mathrm{NG}},s^{\mathrm{NG}}\right]\rangle_{\mathrm{c}}\,,
    \end{equation}
    which is identical to the `trispectrum bias'.

\begin{figure}
        \centering
        \includegraphics[width=\columnwidth]{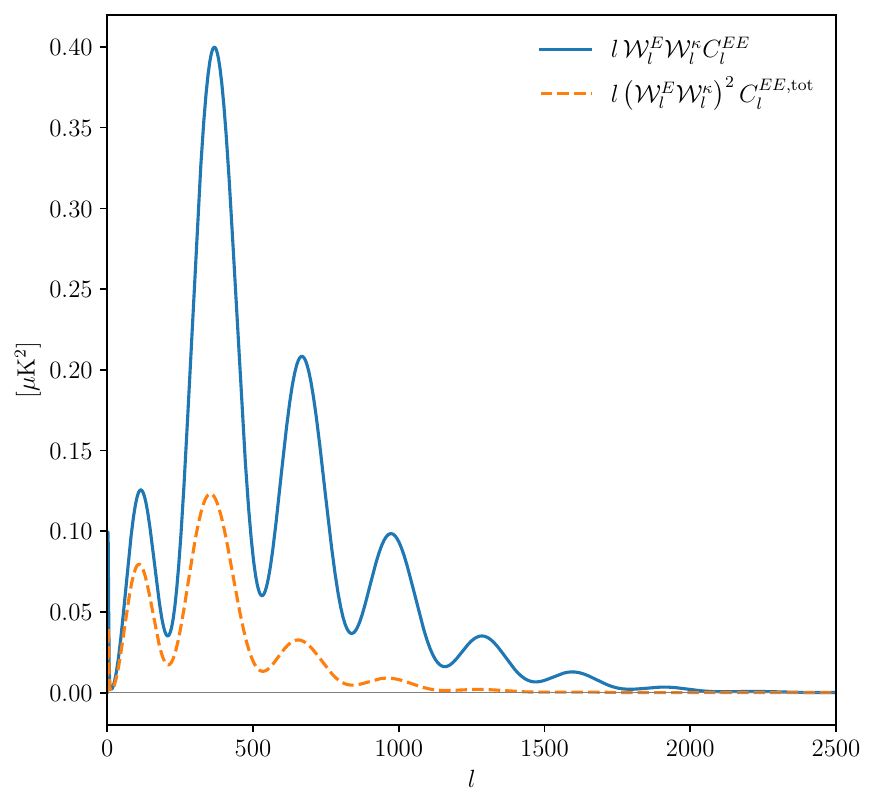}
        \caption{Integration kernels in equations~\eqref{eqn:cross_theory} (blue) and~\eqref{eqn:auto_theory} (orange), highlighting what scales of the lensing convergence are most important when delensing large-scale $B$-modes.}
        \label{fig:integrand_kernels}
    \end{figure}

    These biases to lensing reconstructions have been studied extensively before and are known to be at the level of several percent if unmitigated. It is therefore pertinent to explore what impact they can have on the delensing procedure. In subsequent sections, we will use simulations to do precisely this. However, we can gain some preliminary intuition by investigating what angular-size lenses are most relevant for delensing, and seeing what the $\Delta C^{\hat{\kappa} \hat{\kappa}}$ and $\Delta C^{\kappa \hat{\kappa}}$ biases look like on those scales. Given that we are interested in $B$-modes with $l<300$ and $C^{EE}_{l}$ peaks near $l\sim 1500$, growing rapidly with $l$ (see figure~\ref{fig:cls}), the bulk of the integrals in~\eqref{eqn:full_theory_cross} and~\eqref{eqn:full_theory_auto} will come from the region in multipole-space where $l\ll l'$. In this limit, the integrands simplify extensively, giving
    \begin{align}\label{eqn:cross_theory}
         \Delta C_l^{\tilde{B}\times \hat{B}^{\mathrm{lens}}} \sim \frac{1}{4\pi} \int d l' l' \mathcal{W}^{E}_{l'} \mathcal{W}^{\kappa}_{l'} C^{EE}_{l'}
        \Delta C^{\kappa \hat{\kappa}}_{l'}\,,
    \end{align}
    and
    \begin{align}\label{eqn:auto_theory}
        \Delta C_l^{\hat{B}^{\mathrm{lens}}\times \hat{B}^{\mathrm{lens}}} \sim \frac{1}{4\pi} \int d l' l' \left(\mathcal{W}^{E}_{l'} \mathcal{W}^{\kappa}_{l'}\right)^2 C^{EE, \mathrm{tot}}_{l'}
        \Delta C^{\hat{\kappa} \hat{\kappa}}_{l'}\,,
    \end{align}
    both of which are independent of $l$, as appropriate for the large-scale lensing $B$-modes, which resemble white noise. The integration kernels that accompany $C^{\kappa \hat{\kappa}}$ and $C^{\kappa \hat{\kappa}}$ in the integrands above are plotted in figure~\ref{fig:integrand_kernels}. Notice that modes of $\kappa$ with $L\gsim2000$ are irrelevant when delensing the large-scale $B$-modes; in fact, the majority of the information is coming from $L\lsim1000$.

    This is a important insight. The effect of the bispectrum biases --- equations~\eqref{eqn:bispec_bias}, ~\eqref{eqn:prim_bispec} and~\eqref{eqn:sec_bispec} --- is to suppress power on large scales\footnote{Heuristically, this can be understood as follows. In regions of negative convergence, the CMB power spectrum is shifted to smaller angular scales~\cite{ref:dodelson_lensing_textbook}--- i.e., to the right. From figure~\ref{fig:cls}, it is clear than this results in more power at $l\sim 3000$ relative to the unlensed scenario. This is at the core of how a QE extracts the lensing signal: it interprets any excess of power at large $l$ as evidence of there being below-average $\kappa$ in the region. When the excess power is due to foregrounds, the estimator will return a value of $\hat{\kappa}$ that is biased low.} while adding it on small scales, with the transition between the two regimes happening somewhere in the multipole range $2000\lsim L\lsim2500$, the exact value being experiment-dependent~\cite{ref:van_engelen_15}. The bispectrum biases to lensing spectra are therefore negative across the scales where the integration kernels collect their signal. We thus expect, from equation~\eqref{eqn:cross_theory}, that  $C_l^{\tilde{B}\times \hat{B}^{\mathrm{lens}}}$ will be biased low. And once we take into account the factor of $-2$ preceding it in equation~\eqref{eqn:bias_TT}, we learn that it will contribute a positive bias to the power spectrum of delensed $B$-modes.

    The situation is less clear-cut for $\Delta C_l^{\hat{B}^{\mathrm{lens}}\times \hat{B}^{\mathrm{lens}}}$. Though the primary and secondary bispectrum biases are guaranteed to make a negative contribution, this will be offset to some extent by the trispectrum bias. Being a four-point function of the foreground amplitudes, the trispectrum bias is particularly sensitive to the extent to which individual sources can be removed~\cite{ref:van_engelen_15} --- it most certainly will dominate unless the removal is extensive. This sensitivity to masking, compounded with uncertainties in foreground modeling, makes it difficult to predict the relative amplitude of the different contributions (though analytic work is ongoing~\cite{ref:cosmoblender_in_prep}), and thus the sign of $\Delta C_l^{\hat{B}^{\mathrm{lens}}\times \hat{B}^{\mathrm{lens}}}$. Failing a categorical prediction, we at least expect to see evidence of cancellations between terms in the form of a somewhat reduced bias amplitude.

    For completeness, we note that the $TE$ reconstruction is in principle vulnerable to a bias analogous to the secondary bispectrum bias,
    \begin{align}\label{eqn:bias_TE}
        \Delta C_l^{BB, \mathrm{res}} =& \wick[offset=1.5em]{ h_l \left[ \langle   \c {E}^{\mathrm{obs}} \hat{\kappa}^{\mathrm{TE}} \left[s^{\mathrm{NG}}, \tilde{E} \right] \c {E}^{\mathrm{obs}} \hat{\kappa}^{\mathrm{TE}} \left[s^{\mathrm{NG}}, \tilde{E}\right]\rangle_{\mathrm{c}}\right] }\,.
    \end{align}
    %
    There are also additional contributions from coupling arrangements that feature one or both of the template-$E$-mode legs in the trispectrum; though we omit them here, they are described in appendix~\ref{appendix:TE_biases}. As we shall soon see, the bias terms associated with $TE$ reconstructions are negligibly small.
    
    The breakdown of bias terms in equations~\eqref{eqn:bias_couplings_cross},~\eqref{eqn:bias_couplings_auto} and~\eqref{eqn:bias_TE} is only provided for illustration purposes. In what follows, we will evaluate equation~\eqref{eqn:bias} directly using simulations, thus including all possible contributions.
    
    \section{\label{sec:methods} Computational methods}
    \subsection{\label{sec:sims} Simulations}
    We measure these effects using the Websky simulation~\cite{ref:websky}: a single, full-sky realization of the microwave sky. At the lowest level, it relies on the efficient peak-patch algorithm~\cite{ref:stein_et_al_19} to construct halo catalogues with a mass resolution of $10^{12}\,\mathrm{M}_{\odot}$ at various redshifts, in the cosmology best-fitting the Planck 2018 data~\cite{ref:planck_params_18}. Emission from various processes is then assigned to these halos based on prescriptions from astrophysical models, and this is then projected along the line of sight. The cornerstone of the tSZ model is the `Battaglia' pressure profile of Ref.~\cite{ref:battaglia_et_al_12}, while for the CIB, point-like galaxies are distributed in halos according to the CIB halo model of Ref.~\cite{ref:shang_et_al_12} with the best-fit parameters of Ref.~\cite{ref:hermes_viero_wang}. Testament to the effectiveness of these models is that the tSZ-CIB correlation in the simulation is in good agreement with measurements by Planck~\cite{ref:planck_cib_tsz_correlation}. In addition to the baseline Websky products, we also include the catalogs of radio sources developed by~\cite{ref:li_et_al_21}, which match the realization of the large-scale structure in Websky. Other astrophysical effects, such as CMB lensing or the kinetic SZ effect, are produced by fluctuations on scales so large that they are not bound in halos. In those cases, WebSky projects the emission from the `field' --- calculated using Lagrangian perturbation theory --- in addition to the contribution from halos. Note that we do not consider the integrated Sachs-Wolfe effect~\cite{ref:sachs_wolfe_67}, since this only affects large angular scales of the temperature field which are unimportant for the sake of lensing reconstruction. On small scales, the Rees-Sciama effect~\cite{ref:rees_sciama_68} has been shown to be subdominant to the other extragalactic foregrounds~\cite{ref:ferraro_et_al_22}, so we ignore it as well.

    In later sections, we will often refer to the `unmitigated foreground intensity'. This should be taken to mean the combination of all foreground intensity maps at 143\,GHz --- one of the main `science channels' of SO, where the CMB is brightest relative to the foregrounds. A less noisy lensing reconstruction can be obtained by applying the estimators to an Internal Linear Combination (ILC; see, e.g.,~\cite{ref:tegmark_et_al_03}) of observations at different frequencies, instead of just the 143\,GHz map. To approximate this ILC procedure, we combine mock maps of the extragalactic foregrounds and beam-deconvolved white noise at \{27.3, 41.7, 93.0, 143.0, 225.0, 278.0\}\,GHz, at or near the nominal central frequencies of the observation channels of the large-aperture telescope of SO\footnote{Whenever the SO central frequencies do not match the frequency of the simulated maps, we scale the latter on a pixel-by-pixel basis using the model frequency dependence in~\cite{ref:dunkley_et_al_13}}, for which we assume the `goal' noise and beam properties described in~\cite{ref:SO_science_paper}. We use the publicly-available code \texttt{BasicILC}\footnote{\url{https://github.com/EmmanuelSchaan/BasicILC}} to calculate the harmonic-space ILC weights that minimize the variance of the coadded map. When doing so, we take into account not just CMB, extragalactic foregrounds (before point-source removal) and white noise, but also atmospheric noise and Galactic dust emission, as modeled in~\cite{ref:dunkley_et_al_13}.
    
     We remove point sources from the simulations to the extent that the SO LAT is expected to be able to identify them at 143\,GHz. In the case of radio sources, we simply avoid including those with flux above 5\,mJy when building maps from the catalogs of~\cite{ref:li_et_al_21}. An advantage of doing this at the catalog level is that the number of holes that we need to `drill' in the mask is minimized, simplifying later analyses. For the other foregrounds, for which we have access to map-level simulations rather than the catalogs, we identify point sources detected at $5\sigma$ confidence after applying a matched filter to the maps (see, e.g.,~\cite{ref:haehnelt_tegmark_96}). We then set to zero all pixels located within a circle of radius $3'$ around the point source; this entails masking a mere $0.13\%$ of pixels due to tSZ clusters, and $0.33\%$ because of CIB point sources, which suggests that the bias due to lensing-mask correlations is likely to be negligible~\cite{ref:fabbian_et_al_21}. Doing this to the foreground maps, as opposed to the combination of all components, has the benefit that the components not being masked `inpaint' the hole once all maps are combined, thus avoiding spurious artifacts in the lensing reconstruction~\cite{ref:benoit_levy_13}. It is worth noting that it might be possible to achieve more extensive removal of clusters --- and thus tSZ emission --- by searching for sources at the level of the Compton-$y$ maps instead of the individual frequency channels. Similarly, radio sources are more easily detected at frequencies lower than 143\,GHz, and dusty star-forming galaxies at frequencies higher than that. Since the lensing biases are known to be highly sensitive to the extent of point source removal\footnote{Note however that, as pointed out in~\cite{ref:sailer_et_al_21}, point-source masking reduces the trispectrum bias of lensing by a much larger fraction than the bispectrum biases. This can have the unfortunate consequence of spoiling the low-$L$ cancellation between these terms, leading to an overall increase in bias if masking is too aggressive.}~\cite{ref:van_engelen_et_al_14}, upcoming experiments aiming to mitigate delensing biases should explore the impact of various masking schemes on the biases presented in this work.
    
\begin{figure}
        \centering
        \includegraphics[width=\columnwidth]{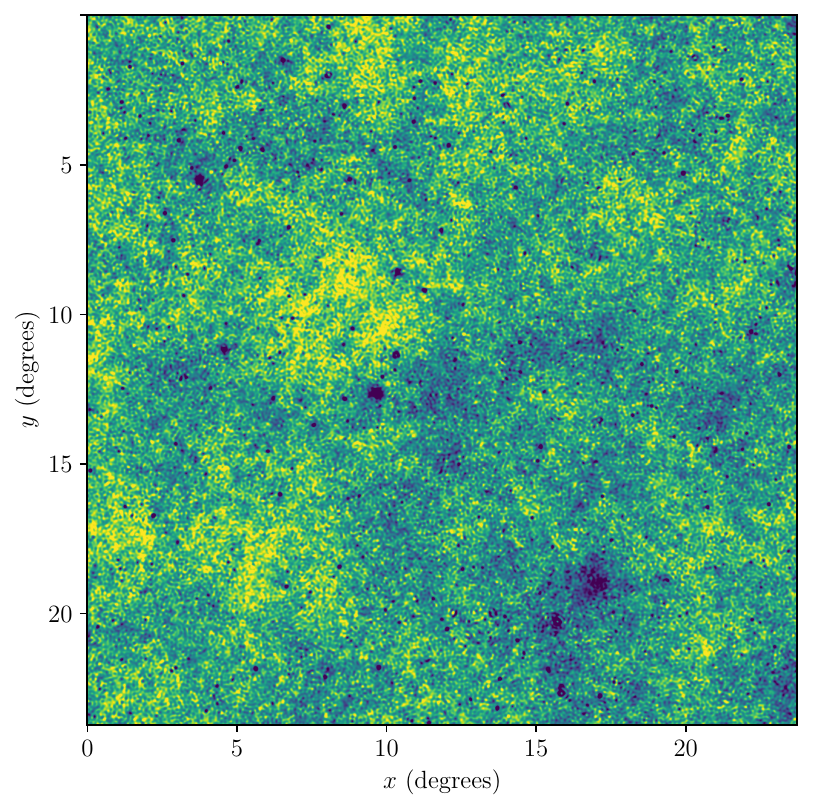}
        \caption{Sample realization of the temperature anisotropies at 143\,GHz produced by the CIB, tSZ and kSZ effects, and radio galaxies.  Point sources detectable with $5\sigma$  confidence in the 143\,GHz channel of the SO LAT have been removed from the individual foreground maps before coadding. The intensity scale ranges from $-25\,\mu$K (dark blue) to $25\,\mu$K (yellow).}
        \label{fig:fg_map}
    \end{figure}

    At this point, we are ready to project the various full-sky emission components onto smaller, flat patches. We do this because publicly-accessible implementations of some of the quadratic estimators we will be testing are only available in the flat-sky limit. From a single full-sky simulation of intensity and polarization with \texttt{Nside}=2048 in the \texttt{HEALPix} pixelization\footnote{\url{http://healpix.sourceforge.net}}, we extract 48 flat, square, non-overlapping patches measuring $23.7\deg$ on a side. These patches are distributed in a homogeneous way across the celestial sphere; in order to avoid excessive distortions when projecting from spherical to planar geometries, we rotate the full-sky maps so that the center of each patch aligns with the origin of equatorial coordinates before projecting it. We pixelize the projected patch into a grid of square pixels, 1\,arcmin on a side, such that there are 1424 of them along each of the two dimensions. A sample realization of the unmitigated foreground component is shown in figure~\ref{fig:fg_map}.

    The projection is applied to both temperature and polarization maps, with the latter requiring a little extra care. In order to avoid $E$-to-$B$ leakage due to finite sky coverage~\cite{ref:lewis_eb_leakage}, we separately project two sets of CMB-only polarization maps: a $QU$ pair which contains only lensed $E$-modes (those from which we will build the lensing template), and another such pair containing only lensed $B$-modes (those that we would like to delens). The resulting, projected fields correctly reproduce the fiducial CMB spectra upon averaging over the 48 patches, up to sample variance.
    
    In parallel, we produce an equal number of Gaussian foreground temperature maps. When simulating the unmitigated foreground component, we do so directly on flat, periodic patches with the same footprint and pixelization properties as those described above. When considering the ILC-cleaned scenario, on the other hand, we produce full-sky realizations at all the relevant frequencies and form the ILC prior to projection, thus reducing the number of expensive projection operations that are required. In either case, we draw harmonic coefficients from a Gaussian distribution with fiducial power spectrum drawn from a smooth fit to the power spectrum measured from the combination of all foregrounds in the full-sky Websky simulation.

    Finally, we combine the CMB-only temperature map with the other components. We take two slightly-different routes depending on whether we are dealing with the case of unmitigated or ILC-cleaned foregrounds. In the former case, we combine the projected CMB-only and foreground-only  temperature maps, convolve them with a Gaussian beam with $\theta_{\mathrm{FWHM}}=1.5$\,arcmin and add white noise to the pixels with a standard deviation of $\Delta_{\mathrm{T}}=6\,\mu\mathrm{K\,arcmin}$. In the foreground-cleaned scenario, on the other hand, the linear combination of all foreground-plus-experiment-noise maps at different frequencies already took place prior to projection, so there is no need to convolve with the beam or add noise again. These steps are the same whether the foregrounds are Gaussian or non-Gaussian; in fact, in order to cancel sample variance, we add the same realization of the noise in both cases.
    
    The reader might have noticed that we do not include atmospheric noise in our simulations. The atmosphere is the dominant source of noise when observing large-scale $l<1000$ temperature anisotropies, so it can increase reconstruction noise --- the reconstructions in~\cite{ref:SO_delensing_paper}, for example, are obtained after discarding $T$ modes below $l\leq500$ --- and degrade delensing efficiency. However, the only way that atmospheric noise can affect the amplitude of the biases we are concerned with in this work is via a relatively-small change to the Wiener filter that is applied to the lensing reconstructions (see section~\ref{sec:templates}), as this modulates the amplitude of the foreground modes. Since characterizations of the atmospheric noise are experiment-dependent and still uncertain, we choose not to factor this effect into our analysis except when calculating the ILC weights (where atmospheric noise does play a leading role at low- and intermediate-$l$). However, we note that our neglect of this contribution will result in slightly more extensive delensing than was forecasted in~\cite{ref:SO_delensing_paper}.
    
    We apply the same beam convolution process to the $Q$ and $U$ maps used to generate the CMB $E$-modes, but add white noise with $\Delta_{\mathrm{P}}=\sqrt{2}\,\Delta_{\mathrm{T}}$. These values are intended to approximate the characteristics of the SO LAT at 145\,GHz (in the `goal' scenario). The $B$-modes are left free of noise to reduce their variance and better isolate the biases of interest.
    
    \subsection{\label{sec:recs} Lensing reconstructions}
    We now explain how our internal reconstructions of CMB lensing are performed. We use the three quadratic estimators presented in section~\ref{sec:qes} --- the standard HO QE, the point-source-hardened QE, and the shear-only QE --- as implemented in the publicly-available code \texttt{symlens}\footnote{\url{https://github.com/simonsobs/symlens}}.
    
    These estimators take as input the projected temperature maps described in the previous section, which contain lensed CMB, foregrounds and experiment noise, possibly after ILC-cleaning. The input fields are inverse-variance-filtered (under the assumption of diagonal covariance) using a smooth fit to the total power in the full-sky simulation, be it the raw map or after foreground-cleaning.
    
    Since we are also interested in measuring the bias to $TE$ reconstructions --- e.g., equation~\eqref{eqn:bias_TE} --- we implement the Hu-Okamoto $TE$ estimator~\cite{ref:hu_okamoto_02}. In this case, the $E$-modes contain CMB and noise, but no foregrounds. We follow~\cite{Planck2018:lensing} and inverse-variance-filter $T$ and $E$ independently, ignoring their correlation.

    The input fields are beam-deconvolved, restricted to a bandpass of $2<l<l_{\mathrm{max}}$ (we consider $l_{\mathrm{max}}=3000,3500$), and masked with a cosine apodization window with a width of 200 pixels ($3.33\deg$). The finite sky coverage gives rise to a mean-field contribution to the reconstructed CMB lensing maps. We estimate it as the mean of the map-level lensing reconstructions, averaged over the entire simulation set; the mean-field subtraction only increases the Gaussian reconstruction noise by (100/$N_{\mathrm{sim}}$)\%~\cite{ref:benoit_levy_13}, where $N_{\mathrm{sim}}$ is the number of realizations used in the computation, in this case 48.
    
\begin{figure*}
        \centering
        \includegraphics[width=\textwidth]{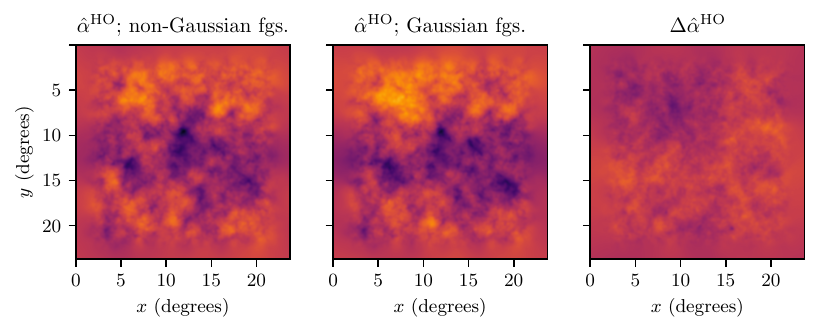}
        \caption{Wiener-filtered magnitude of the deflection angle ($\hat{\alpha}=2\hat{\kappa}/l$) reconstructed using a Hu-Okamoto $TT$ QE applied to mock SO LAT observations at 143\,GHz up to $l_{\mathrm{max}}=3500$ and featuring unmitigated, non-Gaussian (left; same realization as in figure~\ref{fig:fg_map}) or Gaussian (center) extragalactic foregrounds; also shown is the difference between the two (right). The realization of lensed CMB and noise is the same in both pipelines. The intensity scale ranges from $-0.0025$ (dark blue) to $0.0025$ (yellow).}
        \label{fig:kappa_fields}
    \end{figure*}

    In figure~\ref{fig:kappa_fields}, we compare Wiener-filtered reconstructions of the magnitude of the lensing deflection angle ($\hat{\alpha}=2\hat{\kappa}/l$) obtained by applying the standard Hu-Okamoto $TT$ estimator to a simulated patch at 143\,GHz, featuring unmitigated, non-Gaussian (left panel) or Gaussian (center) extragalactic foregrounds. There is a significant difference between the two, as shown in the right-most panel.

    \subsection{\label{sec:templates} Lensing B-mode templates}
    The next step is to construct a lensing $B$-mode template in the style of equation~\eqref{eqn:template}. This takes as input the internal CMB lensing reconstructions obtained in the previous section, and a mock observation of the $E$-modes; the latter comes from projecting the full-sky, lensed-CMB-only Websky $E$-modes onto tiles, and adding experiment noise as explained in section~\ref{sec:sims}.
    
    These input fields must be Wiener-filtered prior to building the template. The $E$-mode filter,
    \begin{equation}
    \mathcal{W}^{E}_l\equiv \frac{\tilde{C}_l^{EE}}{\tilde{C}_l^{EE}+N_l^{EE}}\,,
    \end{equation}
    is calculated from a fiducial noise model that assumes a Gaussian beam and white noise levels appropriate for the SO LAT at 145\,GHz (see section~\ref{sec:sims} and~\cite{ref:SO_science_paper}).

    On the other hand, the Wiener filter for the convergence is
    \begin{equation}\label{eqn:W_phi}
    \mathcal{W}^{\kappa}_L\equiv \frac{C_L^{\kappa\kappa}}{C_l^{\hat{\kappa}\hat{\kappa}}}\,.
    \end{equation}
    We calculate the numerator from the same fiducial model used to generate the Websky simulations. For the denominator, $C_l^{\hat{\kappa}\hat{\kappa}}$,  we simply add to $C_L^{\kappa\kappa}$ an analytic estimate of the Gaussian reconstruction noise, computed from equation~\eqref{eqn:qe_noise}.
    
    We verify that calculating $\mathcal{W}^{\kappa}$ analytically is only marginally less optimal --- in the sense of how tight the error bars on our bias estimates eventually are --- than doing so in a realization-dependent manner, in which $C^{\hat{\kappa}\hat{\kappa}}$ is measured from the simulations. The advantage of the analytic route, however, is that when the filter is the same for both pipelines --- with Gaussian or non-Gaussian foregrounds --- we ensure that, when differencing the delensed $B$-mode power obtained from each, we are not simply picking up effects coming from Gaussian terms calculated with different filtering functions.

    Finally, we restrict both the $E$-mode and convergence fields to angular scales $2<l<3000$, since by the upper end of this range the signal-to-noise is already saturated.
    
    \subsection{Measuring spectra of delensed $B$-modes}\label{sec:measuring_B_spectra}
    Once the lensing $B$-mode template has been built, we can finally delens.  We do this by subtracting the template from noiseless, foreground-free $B$-mode maps, as in equation~\eqref{eqn:Bdel}.

     We apply to the delensed fields a mask that is zero up to 200 pixels from the edges -- avoiding in this way any overlap with the apodized regions of the temperature fields going into the lensing estimator -- and transitions to one over the next 200 pixels in a smooth way, following a cosine curve.
     
\begin{figure*}
        \centering
        \includegraphics[width=\textwidth]{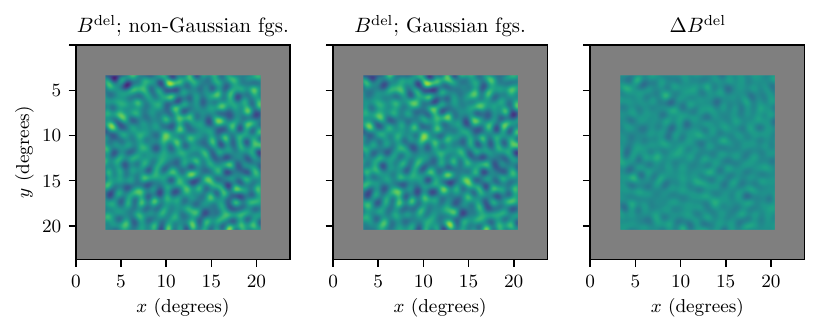}
        \caption{Real-space scalar fields associated with B-modes after delensing noiseless, foreground-free `observations' using the reconstructions in figure~\ref{fig:kappa_fields} --- that is, reconstructions obtained using a Hu-Okamoto QE applied to temperature fields up to $l_{\mathrm{max}}=3500$ featuring unmitigated, non-Gaussian (left) or Gaussian (center) extragalactic foregrounds. The difference between the two (right) has noticeable structure on degree-scales --- modes above $l>300$ have been removed to highlight the degree-scale pattern of interest to primordial $B$-mode searches. Recall that the realization of lensed CMB and noise is the same in both pipelines. Note also that the shaded region corresponds to pixels that are set to zero when measuring spectra. The intensity scale is the same across panels, ranging from $-0.5\,\mu\text{K}$ (dark blue) to $0.5\,\mu\text{K}$ (yellow).}
        \label{fig:Bdel_fields}
    \end{figure*}

    A sample realization of delensed $B$-mode maps is shown in figure~\ref{fig:Bdel_fields}. The panel on the left is obtained using a HO $TT$ estimator applied to 143\,GHz temperature maps featuring non-Gaussian foregrounds, while the foregrounds are Gaussian for the panel in the center. The differences between the two are significant and become stark in the right-most panel, where we restrict the comparison to $l<300$, the scales of interest to primordial $B$-mode searches.

    Since our observations cover only a fraction of the sky, they can suffer $B$-to-$E$ leakage in polarization\footnote{As discussed in section~\ref{sec:sims}, our lensing $B$-modes are measured on tiles extracted from full-sky $QU$ maps containing only $B$-modes (in the full sky, there is no ambiguity between $E$ and $B$, so we can separate them perfectly). On the other hand, our lensing template contains only $B$-modes by construction. There are therefore no $E$-modes in the unmasked sky in our setup, so we are susceptible to $B$-to-$E$ rather than $E$-to-$B$ leakage.}. To circumvent this challenge, we extract pure $B$-modes using \texttt{NaMaster}~\cite{ref:namaster}. We also use this code to calculate the pseudo-$C_l$'s of the delensed field, and deconvolve the mode-coupling induced by the mask. We verify that when the foregrounds are Gaussian the power spectrum of delensed $B$-modes is well modeled by equation~\eqref{eqn:clbbdel_theory}, up to sample variance. We also ensure that the measured bandpowers scatter consistently with the cosmic variance expected of Gaussian fields with their same power spectrum, sky footprint and binning scheme. Note, however, that in the next section we will be able to beat much of this cosmic variance --- and better infer the delensing bias --- by harnessing the fact that the CMB and experiment noise realizations are the same for a given patch and estimator.
    
    \section{\label{sec:results} Results}
    \subsection{Bias when delensing with \\ temperature-based reconstructions}\label{sec:results_single_tracer}
    The machinery described above allows us to evaluate equation~\eqref{eqn:bias} directly from simulations, thus isolating the bias to the power spectrum of residual lensing $B$-modes, $\Delta C_{l}^{BB, \mathrm{res}}$, that ensues after delensing. This contribution will go unmodeled in any analysis of the data that ignores the foreground non-Gaussianity, misleading parameter constraints obtained from $B$-mode spectra.
    
\begin{figure*}
    \subfloat[$l_{\mathrm{max}}=3500$\label{fig:ps_bias_3500}]{%
    \includegraphics[width=0.5\textwidth]{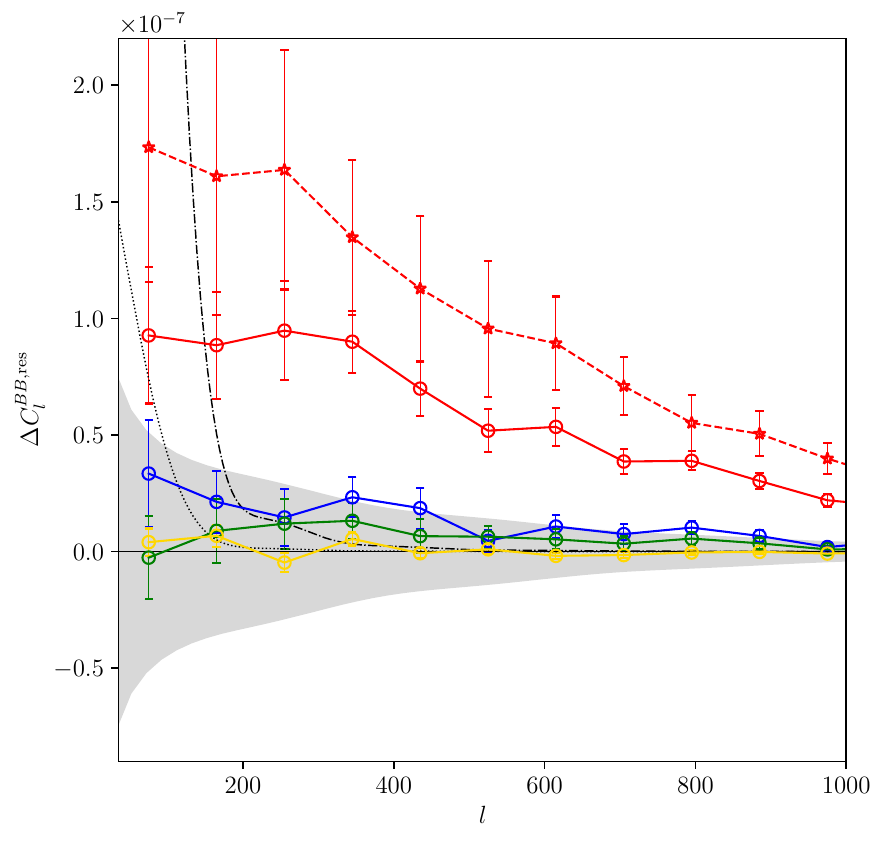}%
    }\hspace*{\fill}%
    \subfloat[$l_{\mathrm{max}}=3000$ \label{fig:ps_bias_3000}]{%
      \includegraphics[width=0.5\textwidth]{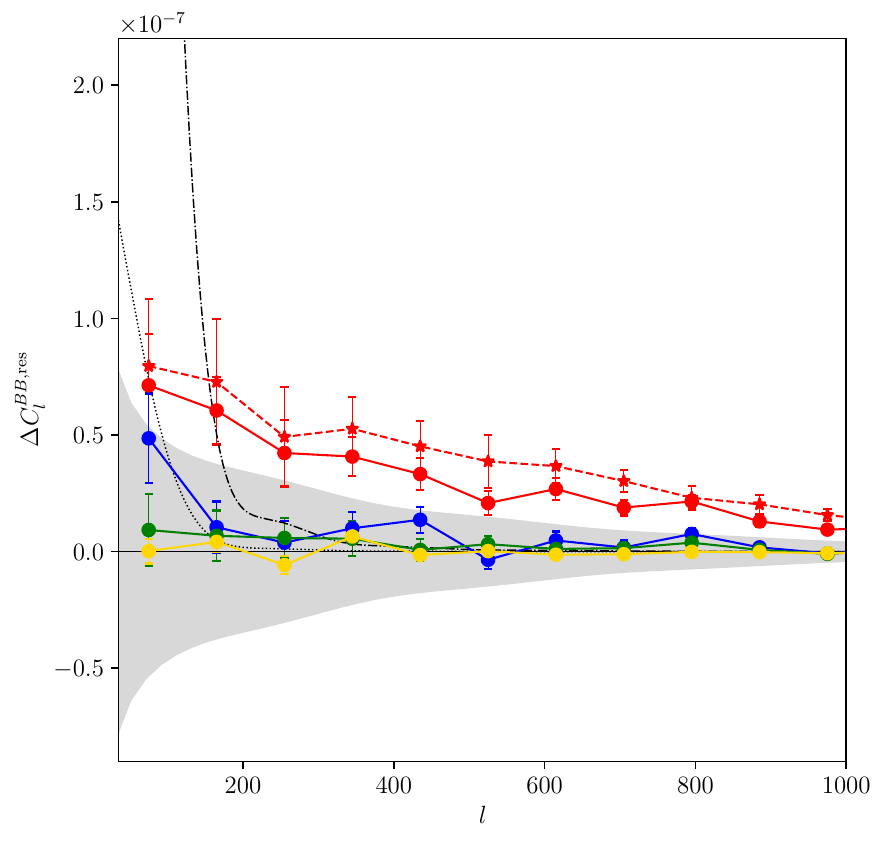}}
      \caption{Bias to the power spectrum of $B$-modes after delensing with various lensing reconstructions obtained from an experiment similar to SO. We compare the Hu-Okamoto $TT$ (red), $TE$ (yellow), point-source-hardened $TT$ (blue) and shear-only $TT$ (green) quadratic estimators. In the left panel, reconstructions are obtained from CMB modes up to $l_{\mathrm{max}}=3500$, and up to $l_{\mathrm{max}}=3000$ in the right. These input maps either include unmitigated foregrounds as they appear at 143\,GHz (solid curves), or they come from a minimum-variance temperature ILC (dashed). In all cases, point sources detected with $5\sigma$ confidence at 143\,GHz are removed. The curves show the mean bias calculated from 48 patches, each covering $O(1\%)$ of the sky, while the error bars show $\pm 1\,\sigma$ of the sample mean. (We combine the measurements between $20<l<1200$ into bins of width $\Delta l =70$.) For comparison, the dashed (dotted) curve shows a tensor contribution with $r=0.01$ ($r=0.001$).  The shaded region denotes the $\pm 1\,\sigma$ fluctuation interval expected of Gaussian fields with the same power spectrum as the residual lensing $B$-modes leftover after delensing with the Hu-Okamoto $TT$ QE (as predicted by theory in the limit of unmitigated Gaussian foregrounds), when observed over 10\% of the sky to approximate the SO SAT footprint, and under the same binning scheme as above.} \label{fig:ps_bias}
    \end{figure*}

    Before we quote any results, let us briefly explain how we will translate modeling inaccuracies to bias on the tensor-to-scalar ratio, $r$. To do this, we will use~\cite{ref:roy_et_al_21}
    \begin{align}\label{eqn:r_shift}
    \Delta \hat{r} = & \left( \sum_{l_{i}} \frac{\left[C_{l_{i}}^{BB}(r=1)\right]^2}{ \mathrm{Var}\left(C_{l_{i}}^{BB,\mathrm{del}}\right)}\right)^{-1} \nonumber \\
    &\times \sum_{l_{i}} \frac{C_{l_{i}}^{BB,\mathrm{unmodeled}} C_{l_{i}}^{BB}(r=1)}{ \mathrm{Var}\left(C_{l_{i}}^{BB,\mathrm{del}}\right)}\,,
    \end{align}
    derived from the maximum-likelihood expression for $\hat{r}$. Here, $l_i$ indexes the $i$-th bin; we employ three bins in the range $30\leq l\leq300$, matching~\cite{ref:SO_science_paper}, with $\Delta l=90$ --- the first three bins in figure~\ref{fig:ps_bias}. $\mathrm{Var}(C_{l_{i}}^{BB,\mathrm{del}})$ is the variance of the binned power spectrum of delensed-$B$-modes (ignoring polarized foregrounds, but including a primordial component), for a sky fraction of 10\%, similar to that covered by the SO SATs; we approximate this variance by its Gaussian component. By setting $C_{l_{i}}^{BB,\mathrm{unmodeled}}=\Delta C_{l}^{BB, \mathrm{res}}$, we will be able to estimate the $\Delta r$ shift caused by the foreground non-Gaussianity as a function of the true value of $r$.

    Figure~\ref{fig:ps_bias_3500} shows the mean bias associated with each of the lensing estimators we consider, along with the standard deviation on this sample mean computed from the scatter of the simulations. For comparison, the shaded gray region shows the $\pm 1$-$\sigma$ fluctuation interval expected of the residual lensing $B$-modes and SAT polarization noise of SO (taking both of these components to be Gaussian) after delensing with an unbiased HO $TT$ QE reconstruction obtained from 143\,GHz maps.

\begin{figure}
        \centering
        \includegraphics[width=\columnwidth]{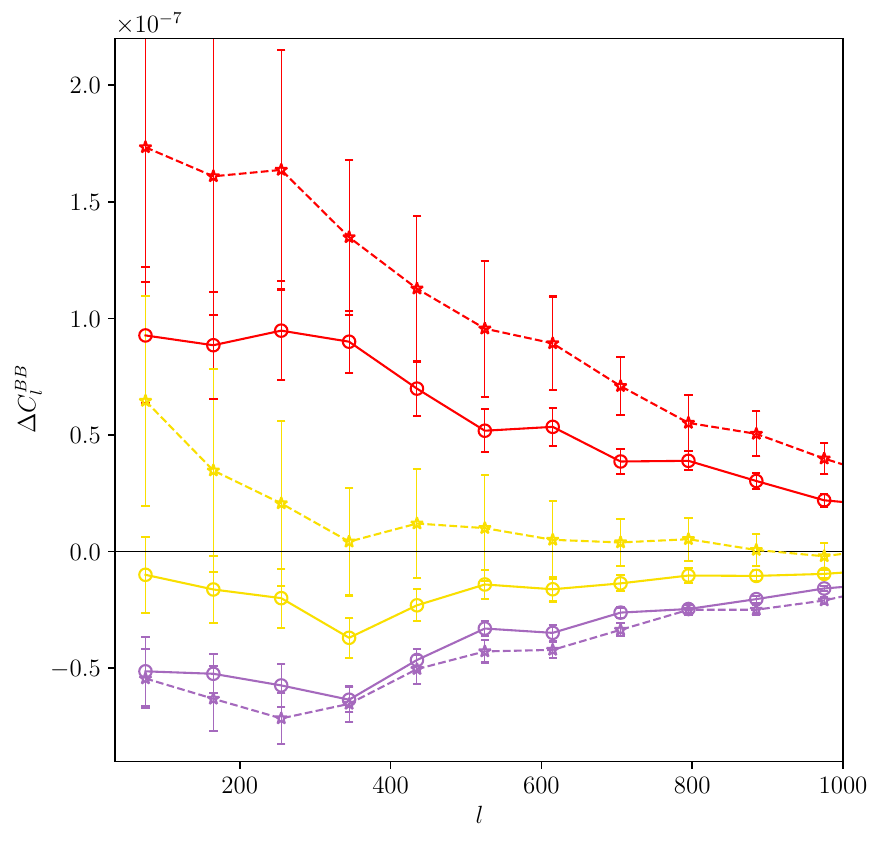}
        \caption{Breakdown of contributions to the $B$-mode power spectrum bias when delensing with a HO $TT$ QE applied to 143\,GHz maps (solid) or an ILC of SO-LAT-like observations (dashed), using CMB modes $l<3500$. The cross-correlation between template and lensing $B$-modes, $C^{\tilde{B} \times \hat{B}^{\mathrm{lens}}}$, is always biased low (magenta), whereas the bias to the template auto-spectrum, $C^{\hat{B}^{\mathrm{lens}}\times \hat{B}^{\mathrm{lens}}}$, (yellow) can be negative when the bispectrum biases dominate on the relevant scales of $C^{\hat{\kappa}\hat{\kappa}}$, or positive when the trispectrum term does instead, as is the case for the ILC scenario. In each case, subtracting twice the magenta curve from the yellow gives the bias to the power spectrum of delensed $B$-modes (red), and we recover the corresponding curve in figure~\ref{fig:ps_bias_3500}.}
        \label{fig:breakdown}
    \end{figure}

    We learn that when delensing with a HO $TT$ QE that takes in temperature modes up to $l_{\mathrm{max}}=3500$ measured at 143\,GHz, the power spectrum is biased high by $\sim$6-7\% at $l<300$. In order to understand the source of these biases, we plot in figure~\ref{fig:breakdown} the shifts to $C^{\tilde{B} \times \hat{B}^{\mathrm{lens}}}$ and $ C^{\hat{B}^{\mathrm{lens}}\times \hat{B}^{\mathrm{lens}}}$ individually\footnote{This might in fact be the more relevant presentation for delensing pipelines based on cross-spectral approaches (e.g.,~\cite{ref:bicep_delensing, ref:SO_delensing_paper})}, before they combine via equation~\eqref{eqn:bias_TT} to give $\Delta C_l^{BB, \mathrm{res}} $. As expected from the arguments in section~\ref{sec:bias_explanation_singletr}, $\Delta C^{\tilde{B} \times \hat{B}^{\mathrm{lens}}}$, which can only be produced by the bispectrum bias to $\Delta C^{\hat{\kappa}\kappa}$, is negative, so it contributes a positive bias to the delensed $B$-mode spectrum. In parallel, $ C^{\hat{B}^{\mathrm{lens}}\times \hat{B}^{\mathrm{lens}}}$ is also biased low, albeit by a smaller amount --- this suggests that the bispectrum biases to $\Delta C^{\hat{\kappa}\hat{\kappa}}$ are dominating over the trispectrum bias on the scales relevant for delensing. Conveniently, this means that the bispectrum contributions to $\Delta C^{\tilde{B} \times \hat{B}^{\mathrm{lens}}}$ and $ \Delta C^{\hat{B}^{\mathrm{lens}}\times \hat{B}^{\mathrm{lens}}}$ cancel each other out partially in the calculation of $\Delta C_l^{BB, \mathrm{res}} $.

\begin{figure*}
    \subfloat[$l_{\mathrm{max}}=3500$\label{fig:r_bias_3500}]{
    \includegraphics[width=0.5\textwidth]{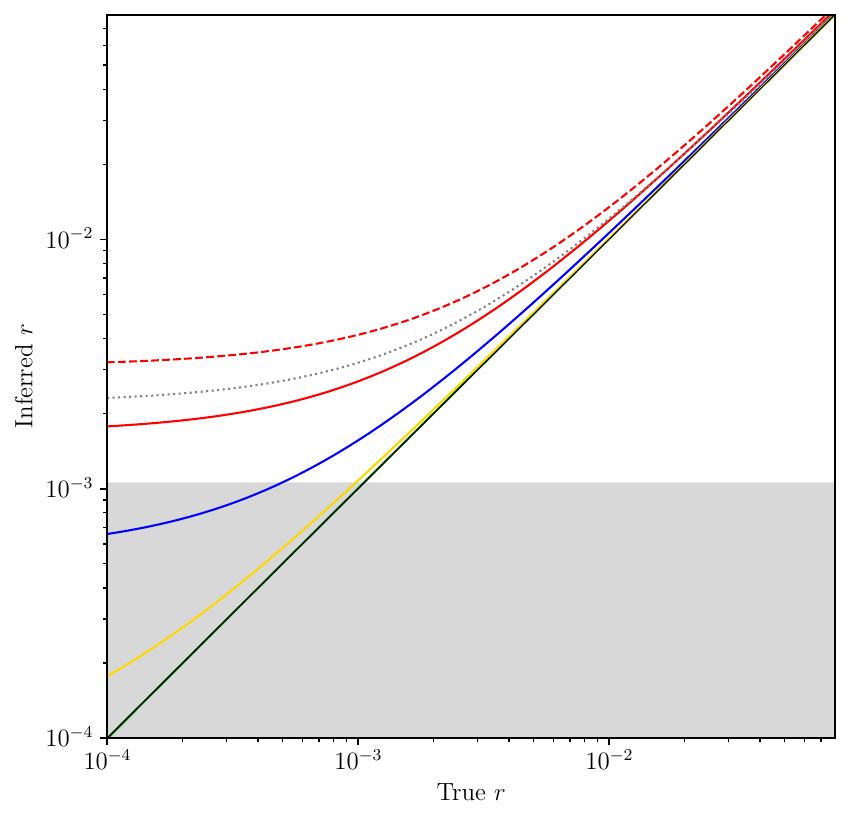}
    }\hspace*{\fill}%
    \subfloat[$l_{\mathrm{max}}=3000$\label{fig:r_bias_3000}]{
    \includegraphics[width=0.5\textwidth]{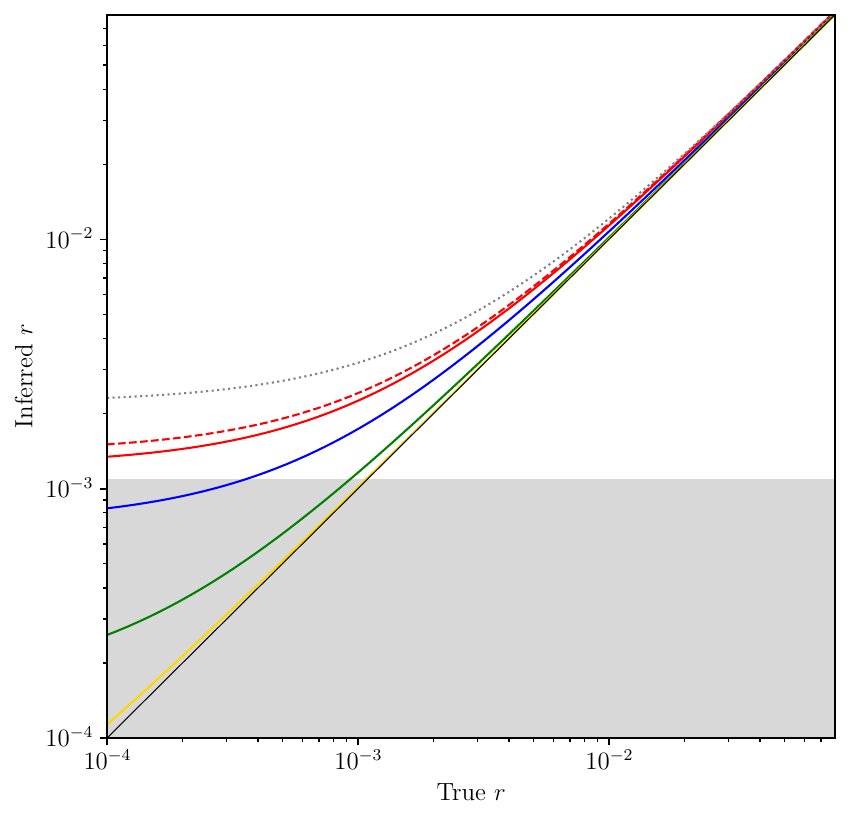}
    }
    \caption{Mean inferred value of $r$ in our simulated patches vs. the true input value, after delensing with the Hu-Okamoto $TT$ (red), Hu-Okamoto $TE$ (yellow), point-source-hardened $TT$ (blue) or shear-only $TT$ (green) quadratic estimators applied to CMB fields featuring non-Gaussian extragalactic foregrounds in temperature as they appear at 143\,GHz (solid) or to an MV-ILC-cleaned map (dashed). Reconstructions are obtained from CMB modes up to $l_{\mathrm{max}}=3500$ (left) or $l_{\mathrm{max}}=3000$ (right). Inferences are biased high for all pipelines, though this is only potentially significant for the Hu-Okamoto estimator: the shaded region shows the $1\,\sigma$ uncertainty afforded by delensing SO data with the Hu-Okamoto TT estimator, including residual lensing $B$-modes and experimental noise, but no foregrounds --- for reference, the dotted line shows the same calculation but featuring also Galactic foregrounds. The biases on $r$ shown here are derived using equation~\eqref{eqn:r_shift}. \label{fig:r_bias}}
    \end{figure*}

    Biases to the power spectrum of residual lensing $B$-modes translate --- via equation~\eqref{eqn:r_shift} --- to biases on $r$: the spurious, additional power leads to a systematic overestimation of the true $r$. This is illustrated in figure~\ref{fig:r_bias_3500}, where we compare the mean inferred values of the tensor-to-scalar ratio to the truth, for various values of $r$. In the null scenario of $r=0$, a naive delensing of SO-like data using a HO $TT$ QE with $l_{\mathrm{max}}=3500$ infers $\hat{r}=1.7\,\times10^{-3}$, $1.5\,\sigma$ away from zero (though any measure of the quality of the fit would likely return a poor value), understanding $\sigma$ to be the fluctuation expected of the $B$-mode spectrum after delensing --- including residual lensing and SAT noise, but ignoring the contribution of polarized foregrounds. Note that our estimate of $\Delta r$ is only the mean value measured from the simulations. This estimate is uncertain due to both sample variance and the simulations being constructed around incomplete astrophysical models of the foregrounds; due to the difficulty in quantifying the latter element, we do not assign error bars to $\Delta r$.

    For completeness, we also investigate the impact of extragalactic foregrounds when delensing with a $TE$ quadratic estimator. In section~\ref{sec:bias_explanation_singletr}, we warned that this pipeline is in principle vulnerable to biases associated with foreground non-Gaussianity in $T$; in particular, we identified a possible secondary bispectrum coupling in equation~\eqref{eqn:bias_TE}. Fortunately, the yellow curves in figures~\ref{fig:ps_bias} and~\ref{fig:r_bias} suggest that for the SO-like experiment we consider the bias is always consistent with zero both at the level of $ C^{BB, \mathrm{res}}$ and of $\hat{r}$.
    
    We now consider four ways of mitigating the delensing bias: 1) multi-frequency cleaning, 2) geometric methods, 3) reducing $l_{\mathrm{max}}$ and 4) modelling it away. Except for the last one, all of these approaches entail a trade-off between lensing reconstruction noise --- and thus delensing efficiency --- and bias. In order to better compare the methods on the basis of this trade-off, we will refer to plots such as figure~\ref{fig:bias_vs_noise}, where we compare $\Delta r$ to $\sigma(r)$ in the null scenario where $r=0$.

    \subsubsection{Multi-frequency cleaning}
    To gauge the impact of foreground cleaning, we consider the case where the HO $TT$ QE reconstruction is derived from a minimum-variance ILC of temperature maps at frequencies near the centers of the nominal channels of the SO LAT; see section~\ref{sec:sims} for details. This multi-frequency combination reduces the map-level noise and consequently allows for more precise lensing reconstructions. Furthermore, Ref.~\cite{ref:sailer_et_al_21} found that it was a useful step in the direction of mitigating biases to the auto- and cross-correlations of temperature-based CMB lensing reconstructions. This is in contrast to what we see here: figure~\ref{fig:ps_bias_3500} shows that the procedure actually worsens the delensing bias relative to the single-frequency scenario described above: $\Delta C_{l<300}^{BB, \mathrm{res}}$ is now at the level of $\sim12\%$ when $l_{\mathrm{max}}=3500$. Naturally, this translates to a larger bias on $r$ --- see the dashed curve in figure~\ref{fig:r_bias_3500} ---, a full $3\sigma$ away from $r=0$ in the null scenario.

     Though this behavior might at first come as a surprise, it is consistent with the understanding we developed in section~\ref{sec:bias_explanation_singletr}, particularly in light of the effect that the ILC is known to have on the individual foreground components and the biases to lensing reconstructions: while the MV ILC suppresses the CIB extensively~\cite{ref:sailer_et_al_21}, it simultaneously boosts the tSZ component in the maps by a factor of $X\sim\sqrt{2}$--$\sqrt{3}$ on the scales relevant to lensing reconstruction (see figure~1 of~\cite{ref:sailer_et_al_21}). The trispectrum is very sensitive to this boost because it scales as the fourth power of the maps, so we expect it to grow by a factor of $X^4$. Since the trispectrum bias is positive and now better able to cancel the negative contributions from the bispectrum biases, $\Delta C^{\hat{\kappa}\hat{\kappa}}$ moves in the positive direction. For the configuration we have chosen, this makes $\Delta C^{\hat{\kappa}\hat{\kappa}}$ smaller than it was in the single-frequency scenario, in agreement with~\cite{ref:sailer_et_al_21}. However, this is actually detrimental to our goals because, as shown by the dashed curves in figure~\ref{fig:breakdown}, a less negative $ \Delta C^{\hat{B}^{\mathrm{lens}}\times \hat{B}^{\mathrm{lens}}}$ spoils the cancellation with $-2\,\Delta C^{\tilde{B} \times \hat{B}^{\mathrm{lens}}}$ that we were seeing in the single-frequency scenario. This means that $\Delta C^{BB, \mathrm{res}}$ is larger when $\hat{\kappa}$ is reconstructed from ILC maps, despite $ \Delta C^{\hat{B}^{\mathrm{lens}}\times \hat{B}^{\mathrm{lens}}}$ being smaller in absolute terms and $\Delta C^{\tilde{B} \times \hat{B}^{\mathrm{lens}}}$ being practically unchanged. Moreover, the scatter in $\Delta C^{BB, \mathrm{res}}$ is larger when working with ILC-cleaned maps: this is because  $ \Delta C^{\hat{B}^{\mathrm{lens}}\times \hat{B}^{\mathrm{lens}}}$ and $\Delta C^{\tilde{B} \times \hat{B}^{\mathrm{lens}}}$ are correlated (they are both dominated by very similar bispectrum biases) so when they are allowed to cancel out, both the mean and the variance of the bias decrease\footnote{We are grateful to Marius Millea for directing our attention to the question of variance.}.
    
    Given that the worsening in delensing bias when reconstructing from a MV ILC can be attributed to an increase in the tSZ trispectrum, it is tempting to consider an alternative ILC where the tSZ component is explicitly deprojected --- that is, nulled by construction~\cite{ref:remazeilles_contrained_ilc}. Though this possibility should be examined in detail in future work, we suspect it will not be all that advantageous for the case at hand. In addition to increasing the noise, tSZ deprojection is known to boost the CIB component by a factor of $\sim \sqrt{10}$ at the map level~\cite{ref:sailer_et_al_21}. This could significantly increase the bispectrum bias, to which we are very sensitive via $-2\,\Delta C^{\tilde{B} \times \hat{B}^{\mathrm{lens}}}$ --- despite this partially cancelling with the bispectrum bias to $C^{\hat{B}^{\mathrm{lens}} \times \hat{B}^{\mathrm{lens}}}$, the latter will grow more slowly. On the other hand, a joint deprojection of both tSZ and CIB is likely to incur too large a penalty in terms of reconstruction noise~\cite{ref:sailer_et_al_21} and delensing efficiency. The optimal solution might result from a compromise between variance reduction and bias mitigation~\cite{ref:Abylkairov_et_al_20, ref:sailer_et_al_21}, or from a combination of multi-frequency cleaning with geometric methods~\cite{ref:darwish_et_al_21}.

    \subsubsection{Geometric methods}
    The class of `geometric methods' comprises quadratic estimators constructed using carefully-chosen weights which differ from those of the standard HO QE and afford the new estimator more immunity to foreground contamination. In section~\ref{sec:qes}, we introduced the point-source-hardened and shear-only estimators, which have proven their worth in mitigating biases to CMB lensing spectra~\cite{ref:schaan_ferraro_18, ref:sailer_et_al, ref:darwish_et_al_21} and appear to be similarly effective against delensing biases.

\begin{figure*}
    \subfloat[$l_{\mathrm{max}}=3500$\label{fig:bias_vs_noise_3500}]{
    \includegraphics[width=0.5\textwidth]{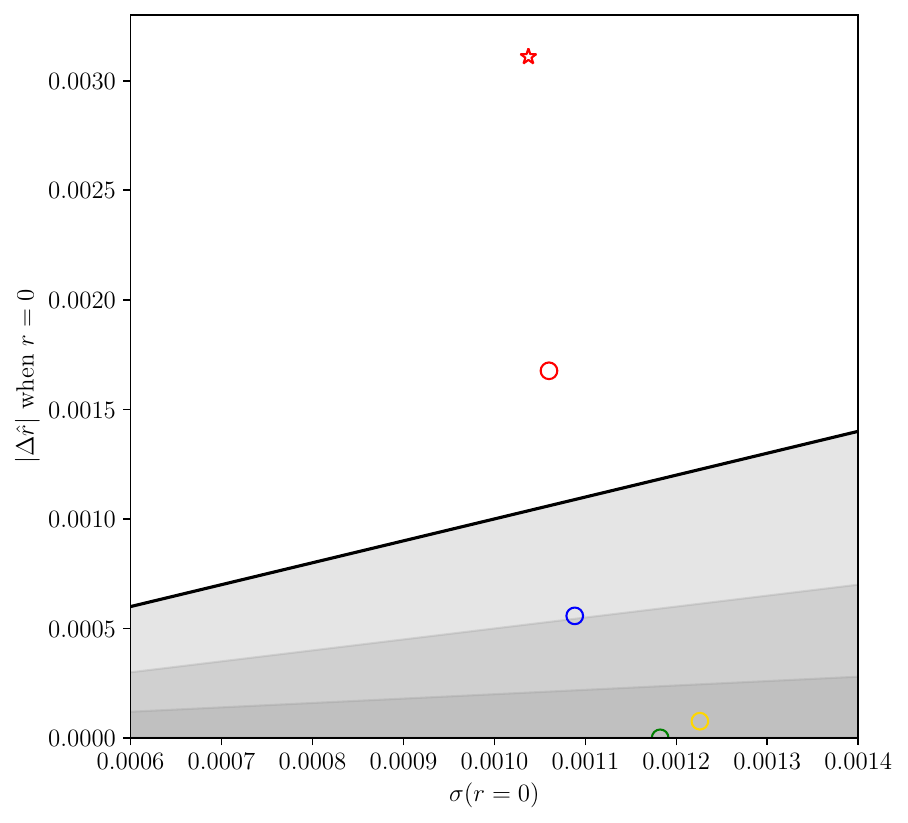}
    }\hspace*{\fill}%
    \subfloat[$l_{\mathrm{max}}=3000$\label{fig:bias_vs_noise_3000}]{
    \includegraphics[width=0.5\textwidth]{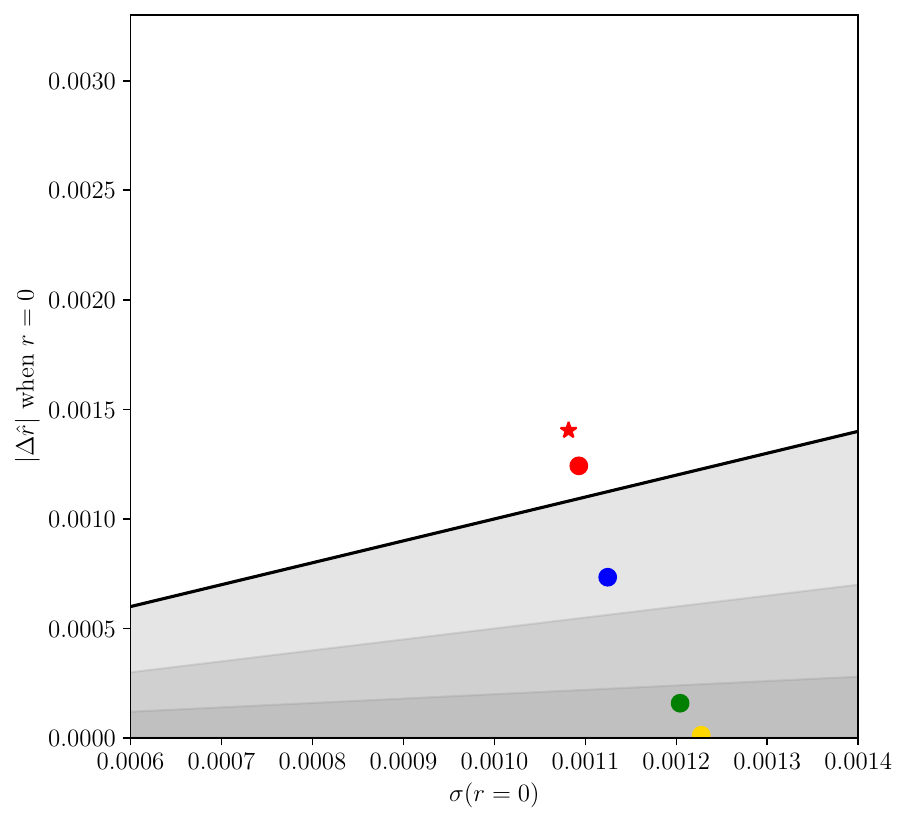}
    }
    \caption{Bias to $r$ vs. statistical uncertainty post-delensing for pipelines involving various quadratic estimators of lensing, in the null scenario where $r=0$. Here, $\sigma(r)$ is  as for the SO SAT's 143 GHz channel white noise levels and approximate sky coverage of $10\%$, with the residual lensing afforded by each estimator, but no polarized foregrounds. The transition between regions of different shading intensity happen at (from bottom to top) $|\Delta \hat{r}|/\sigma^{\mathrm{HO}}(r)=1/5, 1/2, 1$, where $\sigma^{\mathrm{HO}}$ is appropriate for the HO $TT$ QE applied to 143\,GHz maps. The colors are as in figures~\ref{fig:ps_bias} and~\ref{fig:r_bias}, with the difference that the star now denotes the MV-ILC-cleaned pipeline. We see that mitigation strategies based on geometric arguments are most effective at suppressing the bias, while only incurring a moderate degradation in delensing efficiency: they allow us to walk an almost-vertical, downward line on the $\Delta r$-$\sigma(r)$ plane. \label{fig:bias_vs_noise}}
    \end{figure*}

\begin{figure}
    \includegraphics[width=\linewidth]{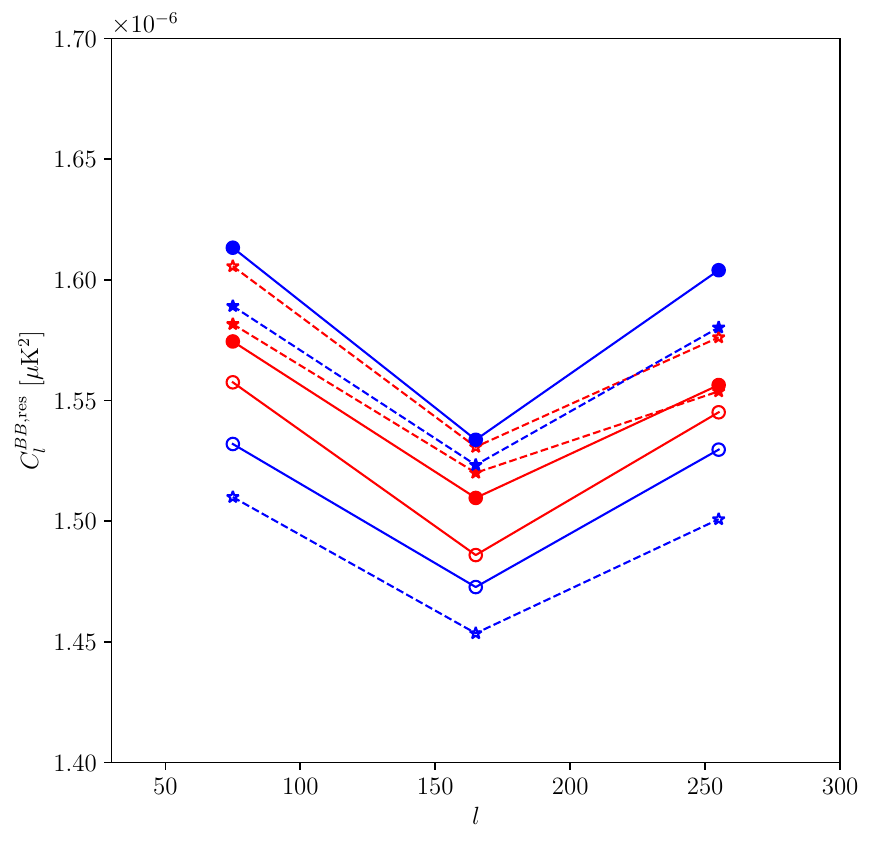}
    \caption{Mean delensed-$B$-mode bandpowers for pipelines involving different QEs applied to simulations featuring non-Gaussian foregrounds. (Note that the delensed spectrum is in fact rather flat, though the restricted $y$-scale does not make it seem that way.) We consider the Hu-Okamoto (red) and point-source-hardened (blue) QEs, applied to 143\,GHz (circle; solid lines) or MV ILC maps (star; dashed lines), using temperature modes $l<3000$ (filled symbol) or $l<3500$ (empty symbol). All things being equal, the HO $TT$ allows for more removal of lensing $B$-modes, but it is also more prone to receiving additional power from the foreground non-Gaussianity. Consequently, among the options we consider, the largest suppression of $B$-mode power is obtained by delensing with a point-source-hardened QE applied to MV ILC maps up to $l_{\mathrm{max}}=3500$.\label{fig:delensing_efficiency}}
    \end{figure}

    The blue curve in figure~\ref{fig:ps_bias_3500} demonstrates that when the delensing pipeline involves a point-source-hardened estimator applied to 143\,GHz temperature maps, $\Delta C^{BB, \mathrm{res}}$ is reduced very substantially relative to the case where the standard HO QE is used. Consequently, the bias to $r$ is also significantly reduced and is now well within the $1\,\sigma$-level, as seen in figure~\ref{fig:r_bias_3500}. Moreover, as shown in figure~\ref{fig:delensing_efficiency}, the total amount of $B$-mode power after delensing is lower than when the HO QE is used: although the PSH reconstruction is slightly noisier (see figure~\ref{fig:QE_noise}) and thus results in somewhat less extensive removal of lensing $B$-modes, this is more than compensated for by the lower amount of spurious power that it receives from foreground non-Gaussianity. All in all, this estimator stands out for its ability to mitigate the bias to acceptable levels at very little cost in terms of delensing efficiency: in figure~\ref{fig:bias_vs_noise_3500}, we see that it reduces $\Delta r$ by a factor of 3 relative to the HO QE, while only degrading $\sigma(r=0)$ by a negligible amount.

    We also consider the shear-only estimator, shown in green in the figures, which is even better at suppressing the bias and comes close to completely neutralizing it. It is actually not surprising that $\Delta C^{BB, \mathrm{res}}$ is not exactly zero despite the foregrounds in the simulation being azimuthally-symmetric. The estimator is only immune to them in the regime where large-scale lenses are reconstructed from CMB fluctuations with much smaller angular sizes, but $B$-mode delensing hinges on relatively small-scale reconstructed lenses, so the separation of scales required for exact immunity is not fully satisfied. At any rate, this estimator is the least-biased out of all the ones we have explored, yielding a $\Delta r$ in figure~\ref{fig:r_bias_3500} that is consistent with zero. The downside is that it comes with greater degradation in delensing efficiency, encapsulated by a larger $\sigma(r)$ in figure~\ref{fig:bias_vs_noise_3500}.

    \subsubsection{Lowering $l_{\mathrm{max}}$}
    Finally, we consider restricting the $l_{\mathrm{max}}$ of the CMB temperature fields from which lensing is reconstructed. This is expected to limit the contamination from extragalactic foregrounds because the latter have a bluer angular spectrum than the CMB --- see figure~\ref{fig:cls} --- exceeding it in amplitude beyond $l\sim3000$ (in the science channels).
    
    Figure~\ref{fig:ps_bias_3000} shows the bias to $\Delta C^{BB, \mathrm{res}}$ for various estimators in the case of $l_{\mathrm{max}}=3000$. Notice that it is reduced in all cases relative to the results in figure~\ref{fig:ps_bias_3500}, for which $l_{\mathrm{max}}=3500$. The same is true for the biases to $r$ in figures~\ref{fig:r_bias_3000} and~\ref{fig:bias_vs_noise_3000}, although it is still above the 1-$\sigma$ level for the standard HO QE, in both the ILC-cleaned and unmitigated scenarios. Since lowering $l_{\mathrm{max}}$ discards information that could otherwise have been used to reconstruct lensing, the delensing efficiency is worsened in all cases, yielding values of $\sigma(r)$ in figure~\ref{fig:bias_vs_noise_3000} higher than what we were seeing in figure~\ref{fig:bias_vs_noise_3500}.
    
    \subsubsection{Modeling}\label{sec:modeling}
    To the extent that the terms in equations~\eqref{eqn:bias_couplings_cross} and~\eqref{eqn:bias_couplings_auto} are the main sources of bias, the effect of the extragalactic foregrounds can be incorporated into models for the power spectrum of delensed $B$-modes --- and the bias removed --- simply by using empirically-calibrated $C^{\kappa \hat{\kappa}}$ and $C^{\hat{\kappa} \hat{\kappa}}$ in equations~\eqref{eqn:full_theory_cross} and~\eqref{eqn:full_theory_auto}. In this approach, $C^{\hat{\kappa} \hat{\kappa}}$ would come from a smooth fit to the lensing reconstruction auto-spectrum, while $C^{\kappa \hat{\kappa}}$ would be the result of applying a similar procedure to the cross-correlation of polarization-only and $TT$ reconstructions\footnote{We are grateful to Anthony Challinor for suggesting this way of obtaining $C^{\kappa \hat{\kappa}}$.}.

\begin{figure}
    \includegraphics[width=\linewidth]{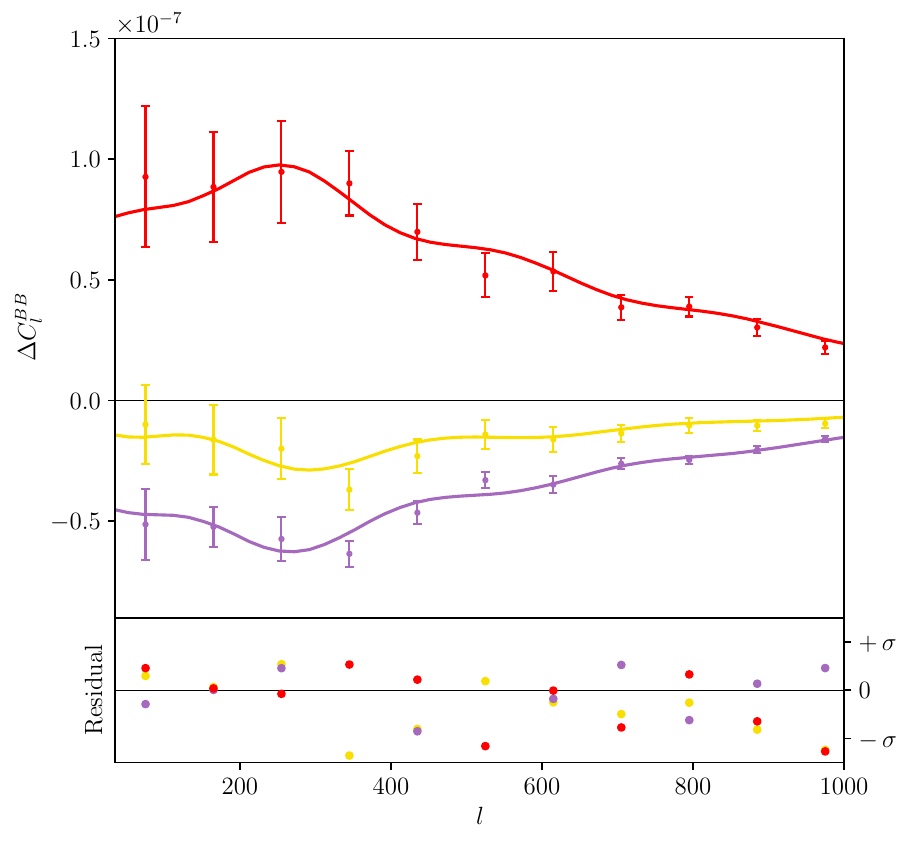}
    \caption{Measured impact of foreground non-Gaussianity on delensed-$B$-mode spectra compared to predictions from empirically-calibrated models. The data points show the mean bias to $C^{\tilde{B} \times \hat{B}^{\mathrm{lens}}}$ (yellow), $ C^{\hat{B}^{\mathrm{lens}}\times \hat{B}^{\mathrm{lens}}}$ (magenta) and $C^{BB, \mathrm{res}}$ (red) arising from a delensing pipeline relying on a HO $TT$ QE applied to unmitigated maps ($l_{\mathrm{max}}=3500$), with error bars denoting $\pm 1\,\sigma$ on the sample mean. (These data are the same as the solid curves in figure~\ref{fig:breakdown}.) The curves, on the other hand, show the difference in model prediction when the $C^{\kappa \hat{\kappa}}$ and $C^{\hat{\kappa} \hat{\kappa}}$ used to evaluate equations~\eqref{eqn:full_theory_cross} and~\eqref{eqn:full_theory_auto} are measured from simulations featuring non-Gaussian versus Gaussian foregrounds. The bottom panel shows the residuals between data and theory when both are binned the same way, normalized to the standard deviation of each datum. This demonstrates that a very accurate model for the delensed data can be built as long as $C^{\kappa \hat{\kappa}}$ and $C^{\hat{\kappa} \hat{\kappa}}$ are determined empirically and are subject to the effects of foreground non-Gaussianity.\label{fig:modelling}}
    \end{figure}

    Figure~\ref{fig:modelling} demonstrates that a model built this way tracks the simulated data very accurately. The solid curves show the difference in model predictions when $C^{\kappa \hat{\kappa}}$ and $C^{\hat{\kappa} \hat{\kappa}}$ are measured from simulations featuring Gaussian versus non-Gaussian foregrounds (we take the average lensing spectra measured across all 48 patches). Notice that the prediction for $\Delta C^{\tilde{B} \times \hat{B}^{\mathrm{lens}}}$, $\Delta  C^{\hat{B}^{\mathrm{lens}}\times \hat{B}^{\mathrm{lens}}}$, and $\Delta C^{BB, \mathrm{res}}$ are all in excellent agreement with the data.  This also implies that the terms we highlighted in equations~\eqref{eqn:bias_couplings_cross} and~\eqref{eqn:bias_couplings_auto} do indeed capture most of the non-Gaussian effects produced by the foregrounds.

    An important benefit of the modeling approach is that it relaxes the requirement to mitigate the bias --- usually at the cost of delensing efficiency --- given that we are now able to model the combination of residual lensing and foreground effects. The focus can then be shifted towards determining what analysis choices minimize the power spectrum of $B$-modes after delensing, whatever its composition in terms of lensing or foreground contributions. These choices include what estimator is used, whether or not a foreground-cleaning procedure is applied, how extensively point sources are masked, what $l_{\mathrm{max}}$ is used for the reconstructions, etc. Unfortunately, the possibility of getting a reduction in power (and thus variance) `for free' is dispelled by the discussion in section~\ref{sec:bias_explanation_singletr}, which suggests that the foregrounds will inevitably add power to the power spectrum of delensed $B$-modes. 
    
    This last point also questions the assumption that delensing with a HO QE will result in lower power than doing so with any other QE. While it is true that, all other things being equal, the HO QE should allow the most extensive removal of lensing $B$-mode power, it is also the one most susceptible to receiving additional power from the foreground non-Gaussianity. Figure~\ref{fig:delensing_efficiency} illustrates the importance of this tradeoff. Among the estimators we consider, the point-source-hardened estimator applied to MV ILC temperature maps up to $l_{\mathrm{max}}=3500$ is in fact the one that ultimately results in the lowest amount of $B$-mode power after delensing. Even though we can model equally well the delensed $B$-mode spectrum resulting from all pipelines, this one is the most powerful one when searching for a primordial component, because of the lower variance.

    \subsection{Bias to a multitracer delensing pipeline}\label{sec:multitracer}\subsubsection{Theory}
\begin{figure}
        \centering
        \includegraphics[width=\columnwidth]{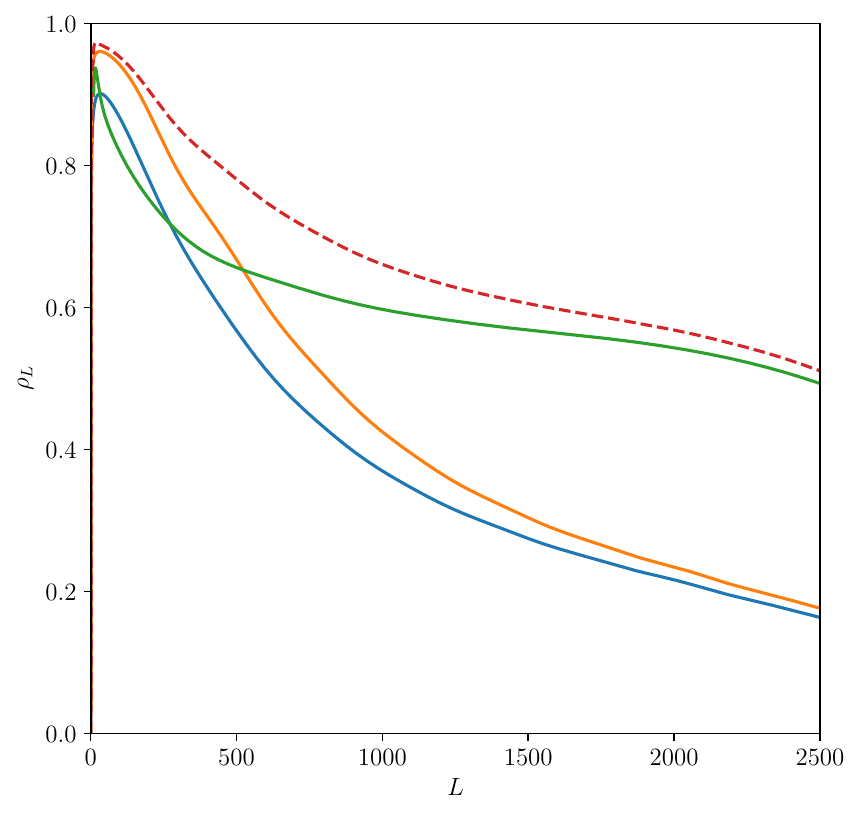}
        \caption{Correlation with CMB lensing of various matter tracers: a HO $TT$ QE (blue), a minimum-variance combination of all the HO QEs (orange), a Planck-like measurement of the CIB at 545\,GHz (green), and the optimal combination of them all (dashed red). All internal reconstructions are derived from CMB modes up to $l_{\mathrm{max}}=3500$. Notice that the co-added tracer is dominated by internal reconstructions on large scales, where the correlation with $\kappa$ is very high, and by the CIB on smaller scales. When combined, the resulting tracer is highly correlated with lensing across the scales that are relevant to large-scale-$B$-mode delensing; cf. figure~\ref{fig:integrand_kernels}.}
        \label{fig:multitracer_rho}
    \end{figure}

    In the previous section, we quantified biases to the delensing procedure when lensing is reconstructed using only $TT$ or $TE$ quadratic estimators. In practice, it is unlikely that these reconstructions will be used in isolation. Instead, the $\hat{\kappa}$ estimate will probably be constructed as a combination of different estimators --- other quadratic combinations of CMB fields~\cite{ref:hu_okamoto_02, ref:maniyar_et_al_21}, such as $EE$ or $EB$, but also maps of tracers of the large-scale structure that are `external' to the CMB~\cite{ref:smith_12_external, ref:sherwin_15, ref:yu_17, ref:manzotti_delensing, ref:karkare_19, ref:SO_delensing_paper}. The advantage of this multitracer approach, illustrated in figure~\ref{fig:multitracer_rho}, is that the co-added tracer maintains a higher degree of correlation with CMB lensing than do any of the tracers individually, and the correlation can be relatively high across the scales of importance to degre-scale-$B$-mode delensing (see figure~\ref{fig:integrand_kernels}). In this section, we investigate what happens to the delensing bias when temperature-based reconstructions are co-added with other tracers of the mass distribution --- as we shall soon see, the problem is not trivial. 

    Be they internal or external, the tracers can all be combined in an optimal manner as $\hat{\kappa}^{\rm comb}_{LM}=\sum_i c^i_L \hat{\kappa}^i_{LM}$, where~\cite{ref:sherwin_15}
    \begin{equation}\label{eqn:multitracer_weights}
        c^i_L = \sum_{j}(\rho^{-1})^{ij}_L \rho^{j\kappa}_L\sqrt{\frac{C_L^{\kappa\kappa}}{C_L^{\hat{\kappa}^i\hat{\kappa}^i}}}
        \,,
    \end{equation}
    Here, $\rho^{i\kappa}_L$ is the cross-correlation coefficient, at multipole $L$, between tracer $\hat{\kappa}^i$ and the true convergence; $\rho^{ij}_L$ is the cross-correlation between tracers $\hat{\kappa}^i$ and $\hat{\kappa}^j$; and $C_L^{\hat{\kappa}^i\hat{\kappa}^i}$ is the angular power spectrum of tracer $\hat{\kappa}^i$. By design, these weights maximize the cross-correlation between the co-added tracer and the true convergence; in the limit of a single tracer, they reduce to the standard Wiener filter in equation~\eqref{eqn:W_phi}.
    
    We will use the weights above to combine $TT$ and $TE$ QEs with external tracers and with other QEs. Ref.~\cite{ref:maniyar_et_al_21} recently pointed out that forming the minimum-variance combination of separate QEs --- the approach of Hu and Okamoto~\cite{ref:hu_okamoto_02}, implicit in equation~\eqref{eqn:multitracer_weights} --- is in principle less optimal than constructing a single, `global minimum variance' (GMV) estimator that finds the combination of pairs of CMB fields that affords the least variance at each scale. Our decision to work with the Hu-Okamoto approach stems from the fact that the difference between methods is expected to be small for SO, and the approach presented here offers more analytic transparency into a problem that is expected to affect the GMV algorithm as well.
    
    Once again, we can use equation~\eqref{eqn:bias} to estimate the delensing bias due to non-Gaussian, extragalactic foreground emission in the observed temperature maps; this time, to a multitracer pipeline. Assuming the primordial CMB is Gaussian, the bias terms are, schematically,
    \begin{widetext}
        \begin{align} \label{eqn:bias_multitracer}
            \Delta C_l^{BB, \mathrm{res}} =& -2 g_l\left[ c^{TT}\langle \tilde{B} \tilde{E} \hat{\kappa}^{TT}[f_{T}+s^{\mathrm{NG}}, f_{T}+s^{\mathrm{NG}} ] \rangle \right] + h_l \left[ (c^{TT})^2 \langle | E^{\mathrm{obs}} \hat{\kappa}^{TT}[f_{T}+s^{\mathrm{NG}},f_{T}+s^{\mathrm{NG}} ] |^{2} \rangle \right] - (s^{\mathrm{NG}} \rightarrow s^{\mathrm{G}})  \nonumber\\
            & -2 g_l\left[ c^{TE} \langle \tilde{B} \tilde{E} \hat{\kappa}^{TE}[f_{T}+s^{\mathrm{NG}}, f_{E}] \rangle \right] + h_l \left[ (c^{TE})^2 \langle | E^{\mathrm{obs}} \hat{\kappa}^{TE}[f_{T}+s^{\mathrm{NG}}, f_{E} ] |^{2} \rangle \right] - (s^{\mathrm{NG}} \rightarrow s^{\mathrm{G}}) \nonumber\\
            & +2 h_l\left[ c^{TT} c^{TE}\langle E^{\mathrm{obs}}  \hat{\kappa}^{TT}[f_{T}+s^{\mathrm{NG}},f_{T}+s^{\mathrm{NG}} ]  E^{\mathrm{obs}}  \hat{\kappa}^{TE}[f_{T}+s^{\mathrm{NG}}, f_{E} ] \rangle \right] - (s^{\mathrm{NG}} \rightarrow s^{\mathrm{G}}) \nonumber\\
            & + 2  h_l\left[ c^{TT} \langle E^{\mathrm{obs}} \sum_{i \neq TT,TE}c^{i} \hat{\kappa}^{i}  E^{\mathrm{obs}} \hat{\kappa}^{TT}[f_{T}+s^{\mathrm{NG}}, f_{T}+s^{\mathrm{NG}} ] \rangle \right]  - (s^{\mathrm{NG}} \rightarrow s^{\mathrm{G}}) \,.
        \end{align}
    \end{widetext}
    The last term in each line represents a duplicate of the previous terms in the line with non-Gaussian foregrounds replaced by their Gaussian version.
    
    The first line is essentially the $TT$ delensing bias in equation~\eqref{eqn:bias_TT}, diluted by the corresponding multitracer weight $c^{TT}$. Similarly, the second line is the diluted version of the $TE$ bias. The third and fourth lines, on the other hand, are new.
    
    The third line is a cross-term between $TT$ and $TE$ reconstructions which produces a plurality of contributions. The two couplings that are likely to stand out among these in terms of significance are
    \begin{align}\label{eqn:TTTE_couplings}
        \Delta & C_l^{BB, \mathrm{res}} \supset \nonumber \\
        &\supset \wick[offset=1.5em]{ 2\,h_l \left[ c^{TT}c^{TE} \langle   \c {E}^{\mathrm{obs}} \hat{\kappa}^{\mathrm{TT}} \left[s^{\mathrm{NG}}, s^{\mathrm{NG}} \right] \c {E}^{\mathrm{obs}} \hat{\kappa}^{\mathrm{TE}} \left[\tilde{T}, \tilde{E}\right]\rangle_{\mathrm{c}}\right] } \nonumber \\
        &+ \wick[offset=1.5em]{ 4\,h_l \left[ c^{TT}c^{TE} \langle   \c {E}^{\mathrm{obs}} \hat{\kappa}^{\mathrm{TT}} \left[\tilde{T}, s^{\mathrm{NG}} \right] \c {E}^{\mathrm{obs}} \hat{\kappa}^{\mathrm{TE}} \left[s^{\mathrm{NG}}, \tilde{E}\right]\rangle_{\mathrm{c}}\right] } \,,
    \end{align}
    that is, terms where we take the Gaussian contraction of the $E$-modes that appear explicitly in the template. The first term above is similar in nature to the primary bispectrum bias of CMB lensing spectra; the second is more similar to the secondary bispectrum bias.
    
    The fourth line produces contributions such as\footnote{Note that we have assumed that the primordial CMB is Gaussian, in which case there are no similar contributions associated with the $TE$ reconstruction.}
    \begin{align}\label{eqn:prim_bispec_bias_multitracer}
        \Delta C_l^{BB, \mathrm{res}} \supset \wick[offset=1.5em]{ 2  h_l\left[ c^{TT}c^{i} \langle \c {E}^{\mathrm{obs}} \hat{\kappa}^{i} \c {E}^{\mathrm{obs}} \hat{\kappa}^{TT}[s^{\mathrm{NG}}, s^{\mathrm{NG}} ] \rangle_{\mathrm{c}} \right] }\,,
    \end{align}
    where $\hat{\kappa}^{i}$ is any matter tracer except for the $TT$ and $TE$  QEs. Cross-terms like this appear in two qualitatively-different, though highly related scenarios. The first case is when the matter tracer features the foregrounds explicitly; for example, when the CIB or the galaxy positions in an imaging survey are used as tracers of the matter (the galaxy maps can feature star-forming galaxies that also make up the CIB, or `radio galaxies' whose emission reaches the microwave range of the spectrum). When this is the case the term is a function of the foreground bispectrum $\langle s s s \rangle$.

    On the other hand, a contribution will arise even if the tracers are foreground-free, merely due to the fact that tracers and foregrounds are correlated because they both trace the same underlying matter distribution. This will be the case for the $EE$, $EB$, or $TB$ reconstructions. (More complicated, though in principle possible, is the case when the weights used to correct for systematic effects in galaxy surveys have large-scale-structure residuals; see, e.g.~\cite{ref:rodriguez-monroy_et_al_22}). The term is then a function of the mixed lensing-foreground bispectrum $\langle \kappa s s\rangle$, a coupling similar to the primary bispectrum bias described in section~\ref{sec:bias_explanation_singletr}; as we saw there, we expect this contribution to be negative.
    
    Note that the couplings we have highlighted can be modeled in a manner similar to what we described in section~\ref{sec:modeling} for a single tracer. The diluted $TT$ and $TE$ contributions can be modeled following the arguments there, but including the multitracer weights in the relevant integrands. Term~\eqref{eqn:prim_bispec_bias_multitracer} can be calculated by modifying equation~\eqref{eqn:full_theory_auto} to include the multitracer weights $c^{TT}$ and $c^{i}$ in the integrand, and replacing $C^{\hat{\kappa} \hat{\kappa}}$ with (a smoothed version of) the measured cross-spectrum between a $TT$ reconstruction and the tracer in question ---  a CIB map, or perhaps the $EB$, $EE$ or $TB$ QE reconstructions. The same goes for the terms in equation~\eqref{eqn:TTTE_couplings}, though this time the multitracer weights are $c^{TT}$ and $c^{TE}$, and $C^{\hat{\kappa} \hat{\kappa}}$ is calibrated against the measured cross-correlation between $TT$ and $TE$ reconstructions.
    
    Alternatively, terms like equation~\eqref{eqn:prim_bispec_bias_multitracer} above can be mitigated by reducing the overlap in multipole space of the multitracer weights. Though impractical when striving to mitigate the bias from the cross-correlation of $\hat{\kappa}^{TT}$ with other internal reconstructions, this approach is certainly feasible when targeting the correlation with external tracers, which are complementary to internal reconstructions in terms of the scales that they target: while internal reconstructions can only accurately reconstruct the largest lenses, external tracers are good at providing the high-$L$ ones; see figure~\ref{fig:multitracer_rho}. Such cuts have already been explored in the context of CIB delensing, either to reduce biases from Galactic dust~\cite{ref:cib_delensing_biases, ref:yu_17}, or because of mode-loss during component separation~\cite{ref:gnilc}; it appears that the delensing efficiency is not majorly affected even if fairly stringent cuts such as the removal of $L<100$ are put in place. Moreover, it has recently been shown that when internal reconstructions are prioritized in the provision of the largest-angular-scale lenses, any possible residual in the power spectrum of delensed $B$-modes tends to flatten out and can be easily marginalized over~\cite{ref:SO_delensing_paper}.

    \subsubsection{Methods}
    Now that we have laid out the theory of what couplings could potentially bias a multitracer delensing pipeline, we set out to quantify them for an experiment similar to the Simons Observatory, with characteristics described in section~\ref{sec:sims}. In what follows, we coadd internal reconstructions from various QEs with CIB maps intended to mimic Planck's observations, thus testing the impact of all the bias terms in equation~\eqref{eqn:bias_multitracer}.
    
    We carry out the $TT$ and $TE$ reconstructions as described in section~\ref{sec:recs}; on the other hand, and for the sake of simplicity, we approximate the $TB$, $EE$ and $EB$ reconstructions as the sum of the input convergence and noise realizations drawn from a Gaussian distribution with the same power spectrum as the $N^{(0)}$ noise corresponding to that estimator, computed using equation~\eqref{eqn:qe_noise}. In all cases, we set $l_{\mathrm{max}}=3500$, and apply the estimators to the `unmitigated' 143\,GHz maps--- that is, we do not perform any foreground cleaning except for removing point sources. We then take the Websky CIB map at 545\,GHz as an approximation to the CIB map that would be used as a matter tracer for delensing, making sure to add detector noise as appropriate for the equivalent channel of Planck, and removing point sources that fall above the detection limit of 350 mJy by setting the pixels to the mean value of the remaining, unmasked pixels. 

    In order to calculate the multitracer weights of equation~\eqref{eqn:multitracer_weights} and combine these tracers optimally, we need all the auto- and cross-spectra of the tracers involved. We measure the auto-correlation of the noiseless 545\,GHz CIB map, as well as its cross-correlation with the input convergence map, on the full sky. We then add an idealized noise power spectrum to the auto-spectrum (using the instrument characterization described in~\cite{ref:planck_18_legacy}), and fit smooth functions to the resulting auto- and cross-spectra\footnote{Ref.~\cite{ref:yu_17} showed that constructing the multitracer weights from smooth spectra avoids biasing the delensed $B$-mode spectrum.}. The remaining ingredients, once the CIB angular spectra have been measured, are all the auto- and cross-spectra of the internal reconstructions. To simplify matters, we assume that the noise covariance matrix between different estimators is diagonal\footnote{Different HO quadratic estimator do in fact have correlated noise (with the exception of $TT$ and $EB$), but this is below the 10\% level~\cite{ref:hu_okamoto_02}.}, and model their cross-correlation as the true convergence power spectrum; on the other hand, we approximate the individual auto-spectra as a sum of the true convergence power spectrum and the $N^{(0)}$ noise biases\footnote{For the $EB$ estimator, we calculate the reconstruction noise assuming that the $B$-modes going into the estimator are masked below $l<300$; this would be required when delensing at the map level in order to avoid bias due to the overlap in modes with the target $B$-modes~\cite{ref:teng_11, ref:baleato_20_internal}.}. From these spectra, we can also calculate the correlation of the various tracers with CMB lensing; this is shown in figure~\ref{fig:multitracer_rho}.
    
    We then build the template following section~\ref{sec:templates}, subtract it from lensing $B$-mode maps, and measure the power spectrum of the resulting, delensed $B$-modes using the methodology in section~\ref{sec:measuring_B_spectra}. By applying this pipeline to temperature fields featuring either Gaussian or non-Gaussian foregrounds, we can evaluate equation~\eqref{eqn:bias_multitracer} for our coadded tracer and isolate the relevant delensing biases. 
    
    \subsubsection{Results}
\begin{figure}
        \centering
        \includegraphics[width=\columnwidth]{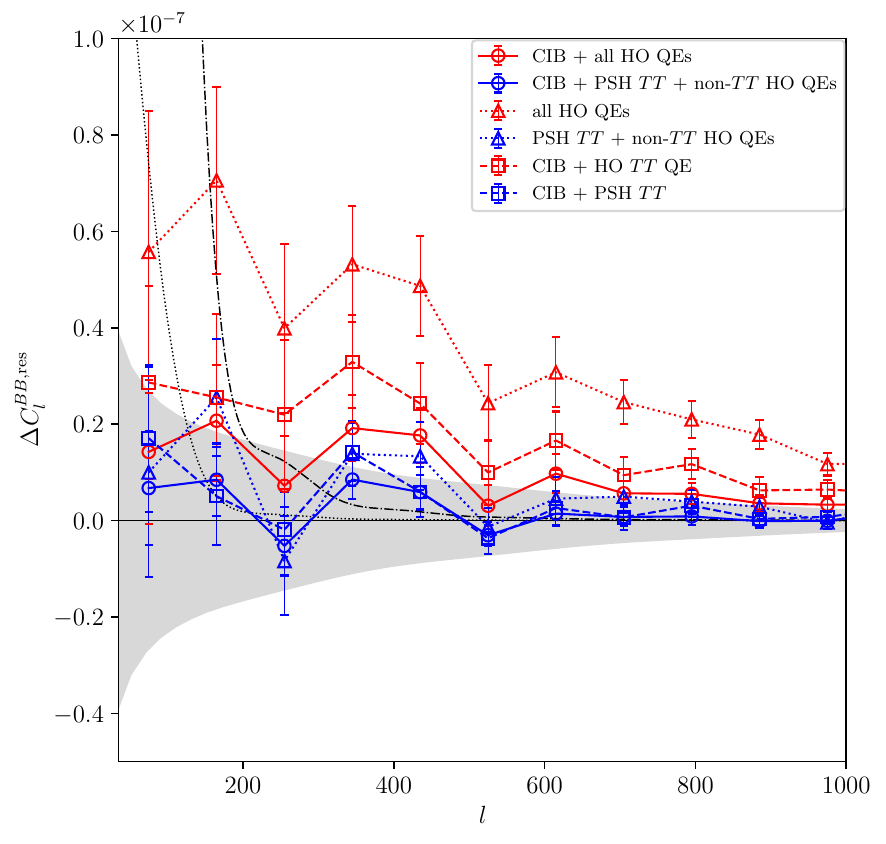}
        \caption{Bias to the power spectrum of $B$-modes after delensing with various multitracer pipelines. (Note that PSH stands for the point-source-hardened QE.) The shaded region denotes the $\pm 1\,\sigma$ fluctuation interval expected of Gaussian fields with the same power spectrum as the residual lensing $B$-modes leftover after delensing with all the Hu-Okamoto QEs and a Planck-like CIB  map at 545\,GHz (as predicted by theory in the limit of unmitigated Gaussian foregrounds), when observed over 10\% of the sky to approximate the SO SAT footprint, and under the same binning scheme as above. Note that the $y$-scale differs from figure~\ref{fig:ps_bias}.}
        \label{fig:ps_bias_multitracer}
    \end{figure}

\begin{figure}
        \centering
        \includegraphics[width=\columnwidth]{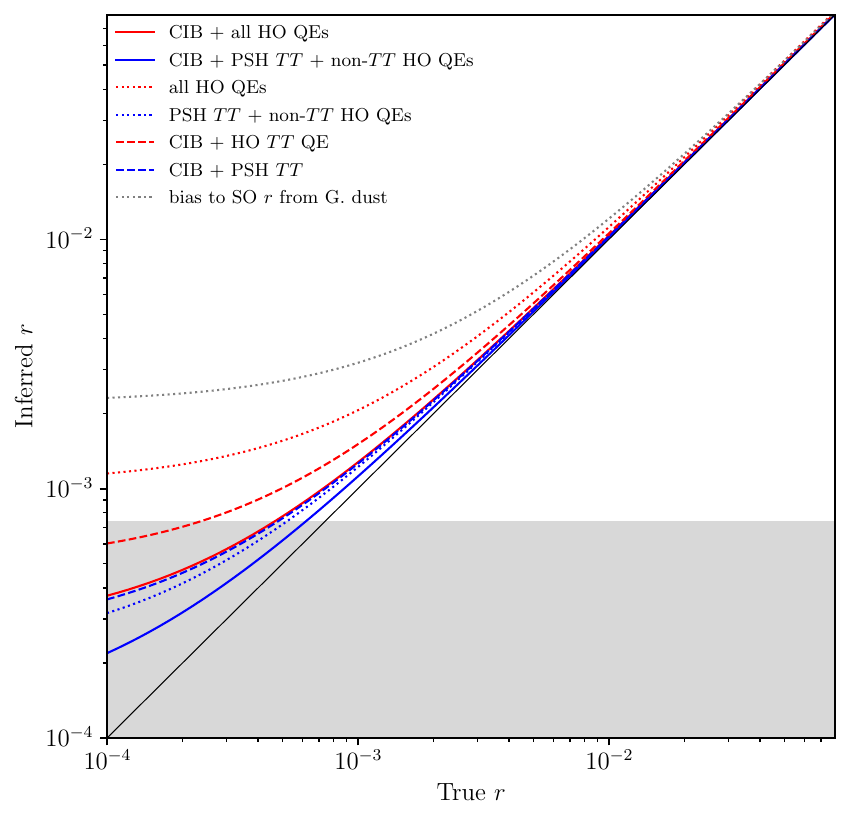}
        \caption{Mean inferred value of $r$ in our simulated patches vs. the true input value, after multitracer delensing. These biases in $r$ are to be attributed to the power spectrum biases in figure~\ref{fig:ps_bias_multitracer}. The shaded region shows the $1\,\sigma$ uncertainty afforded by delensing SO data with a combination of the CIB plus all the HO QEs (including residual lensing $B$-modes and experimental noise, but not foregrounds --- for reference, the dotted black line shows the same calculation but featuring also Galactic foregrounds).}
        \label{fig:r_bias_multitracer}
    \end{figure}

    Figure~\ref{fig:ps_bias_multitracer} shows the bias to the power spectrum of $B$-modes delensed using various multitracer pipelines. These translate to the biases on the inferred value of $r$ shown in figure~\ref{fig:r_bias_multitracer}.

\begin{figure}
        \centering
        \includegraphics[width=\columnwidth]{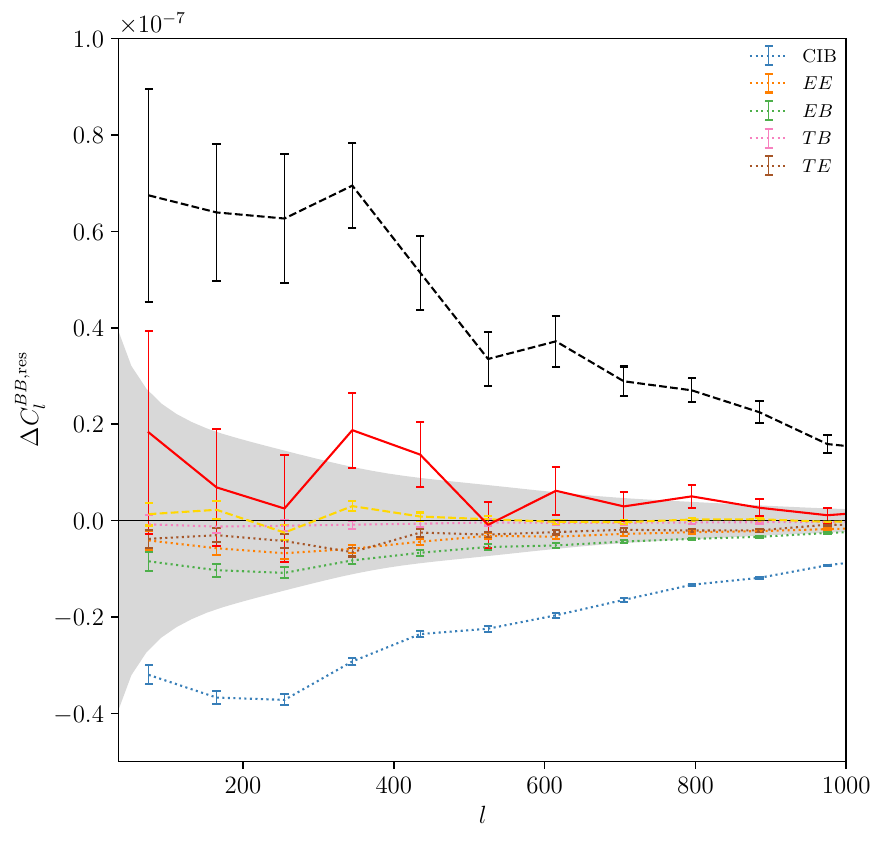}
        \caption{Breakdown of contributions to $\Delta C^{BB,\mathrm{res}}$ when delensing with a combination of $TT$, $TE$, $TB$, $EE$ and $EB$ HO QEs and a Planck-like map of the CIB at 545\,GHz. The shaded region shows the $\pm 1\,\sigma$ (Gaussian) uncertainty due to residual lensing $B$-mode power when observed over 10\% of the sky, and the binning shown. The diluted bias from $TT$ reconstruction (dashed, black) partially cancels with the cross-terms in equation~\eqref{eqn:bias_multitracer} (dotted; the legend shows which tracer is involved) --- especially, with the coupling involving the CIB. Also shown are the $TT$-$TE$ cross-term (dotted, brown) and the diluted $TE$ bias (dashed, yellow). All of these effects combine to give the red solid curve, which matches the total bias (red, solid curve in figure~\ref{fig:ps_bias_multitracer}) up to sample variance.}
        \label{fig:breakdown_ps_bias_multitracer}
    \end{figure}

    The first thing to note is that the hierarchy in bias amplitude of the different pipelines involving the HO $TT$ QE cannot be explained just in terms of the dilution of the $TT$ reconstruction bias. This is a hint that the other couplings in equation~\eqref{eqn:bias_multitracer} are playing a significant role. Take, as a case study, a multitracer pipeline built around a combination of the $\{TT,TE,TB,EE,EB\}$ QEs of Hu and Okamoto, combined with a Planck-like map of the CIB at 545\,GHz. In figure~\ref{fig:breakdown_ps_bias_multitracer}, we dissect the total bias resulting from this pipeline into its various constituent contributions\footnote{We compute the last line in equation~\eqref{eqn:bias_multitracer} directly from equation~\eqref{eqn:prim_bispec_bias_multitracer}, with nothing but foregrounds in the inputs to the QE. This explains the smaller error bars shown by these terms in figure~\ref{fig:breakdown_ps_bias_multitracer} relative to other contributions.} Notice that not only are the cross-terms in equation~\eqref{eqn:prim_bispec_bias_multitracer} significant, but they also partially cancel with the diluted bias from $TT$ reconstruction, yielding a lower bias (red, solid) than would be expected merely on the grounds of dilution (black, dashed). Naturally, the coupling is particularly strong, and the cancellation more extensive, when the CIB is used as an external tracer for delensing, but the effect is qualitatively the same for any other tracer of the matter distribution, including the $EE$, $EB$ and $TB$ reconstructions. The $TT$--$TE$ cross-term, equation~\eqref{eqn:TTTE_couplings} and similar, also contribute to the cancellation, albeit by a smaller amount. 

\begin{figure}
        \centering
        \includegraphics[width=\columnwidth]{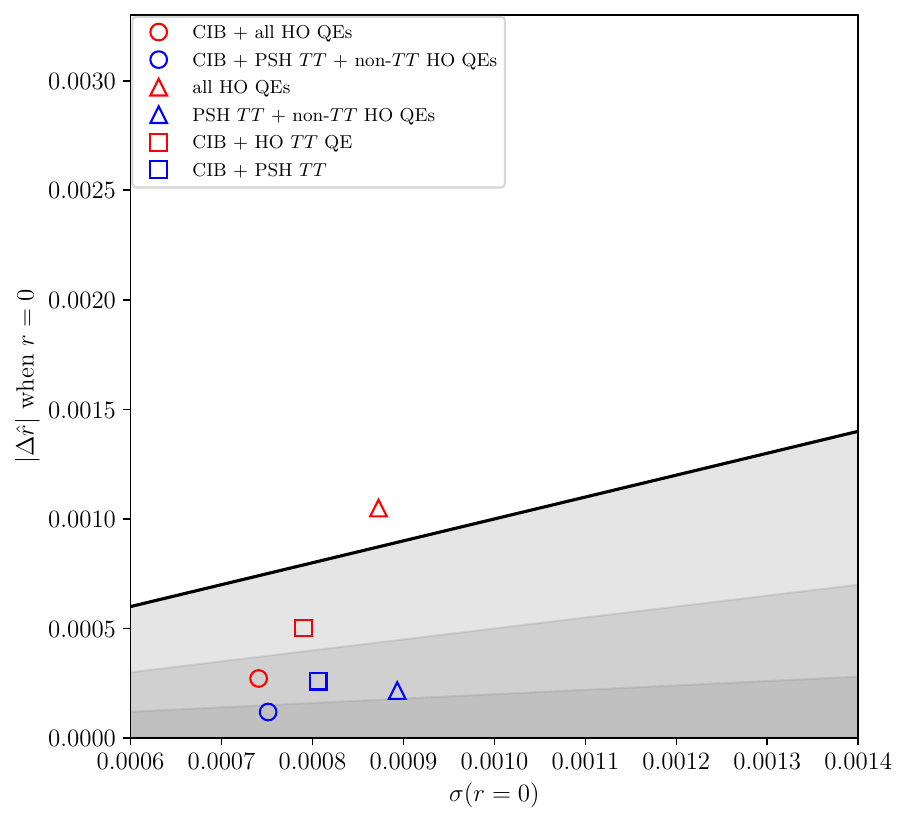}
        \caption{Bias to $r$ vs. statistical uncertainty after delensing with various multitracer pipelines, in the null scenario where $r=0$. Here, $\sigma(r)$ is  as for the SO SAT's 143 GHz channel white noise levels and approximate sky coverage of $10\%$, with the residual lensing afforded by each pipeline, but no polarized foregrounds. The transition between regions of different shading intensity happen at (from bottom to top) $|\Delta \hat{r}|/\sigma_{0}(r)=1/5, 1/2, 1$, where $\sigma_{0}$ is appropriate for the combination of a HO $TT$ QE with all the other HO QEs (applied to 143\,GHz maps) and a Planck-like map of the CIB at 545\,GHz. The pipelines are the same as in figures~\ref{fig:ps_bias_multitracer} and~\ref{fig:r_bias_multitracer}, and the colors and symbols also match.
        }
        \label{fig:bias_vs_noise_multitracer}
    \end{figure}

    These cancellations are good news from the point of view of delensing. They suggest that it might be possible to use the more contaminated --- but more effective --- HO $TT$ QE down to smaller scales (higher $l_{\mathrm{max}}$) than one would be inclined to pursue based on the findings in section~\ref{sec:results_single_tracer}.
     Consider figure~\ref{fig:bias_vs_noise_multitracer}, where we compare $\Delta r$ to $\sigma(r)$ after delensing with various multitracer pipelines, in the null scenario where $r=0$. When the HO $TT$ QE is co-added with all the other HO QEs, the bias is above 1-$\sigma$. Coadding instead with the CIB not only improves the delensing effciency, but it also gives a lower bias on $r$ thanks to the cancellations brought about by a strong $TT$-CIB cross-term from equation~\eqref{eqn:prim_bispec_bias_multitracer}. Finally, when all the HO QEs are combined with the CIB, the delensing efficiency\footnote{Note that throughout this paper we see more extensive delensing than was forecasted in~\cite{ref:SO_delensing_paper}. There are several reasons for this. First, we ignore atmospheric noise, which led~\cite{ref:SO_delensing_paper} to discard temperature modes below $l<500$. Second, we constrain $r$ over the $BB$ range $30<l<300$, matching~\cite{ref:SO_science_paper}, whereas~\cite{ref:SO_delensing_paper} used $50<l<200$. Third, we often report results for $l_{\mathrm{max}}=3500$, while they only ever consider $l_{\mathrm{max}}=3000$. And last, but certainly not least, we assume the SAT observes 10\% of the sky, as in~\cite{ref:SO_science_paper}, whereas~\cite{ref:SO_delensing_paper} works with a smaller patch covering 5\% of the sky.}, the dilution, and the cancelations are all at a maximum (among the cases we consider), and the bias is below $\sigma/2$.

     If needed, the bias could be suppressed even further by resorting to the methods described in section~\ref{sec:results_single_tracer}: reducing $l_{\mathrm{max}}$ for the $TT$ estimator, carefully removing foregrounds (bearing in mind that foreground removal is not guaranteed to improve the situation, as seen above), or adopting alternative, more robust estimators such as those described in section~\ref{sec:qes}. As an example of the latter approach, figure~\ref{fig:bias_vs_noise_multitracer} shows that replacing the HO $TT$ QE with the point-source-hardened estimator reduces the bias while only incurring a very modest when it comes to removing lensing $B$-modes.

    \section{\label{sec:conclusion} Conclusions}
    Delensing is already an essential ingredient in any effort to constrain primordial gravitational waves from the $B$-mode polarization of the CMB. In this work, we characterized, for the first time, a potential source of bias to delensing pipelines: the non-Gaussian nature of extragalactic foregrounds. We show that the effect can be major when delensing hinges on internal reconstructions of lensing derived from the CMB temperature anisotropies, as will be the case for SO. However, it will be relevant more generally, including for the South Pole Observatory and CMB-S4, whenever temperature information is used to improve the quality of the lensing recontruction (as in, e.g.,~\cite{ref:wu_et_al_19}), even if polarization dominates.

    The problem is that the foreground non-Gaussianity induces higher-point contributions that add power to the power spectrum of delensed $B$-modes, potentially confusing inference pipelines that ignore the foregrounds or approximate them as being Gaussian. In particular, the cross-correlation of a lensing $B$-mode template constructed from a standard $TT$ QE with the true $B$-modes is biased low, whereas the auto-spectrum of such a template can be biased either low or high, depending on the relative strength of the bispectrum and trispectrum biases of lensing. When the former dominates, convenient cancellations appear  in the power spectrum of delensed $B$-modes between biases to the template auto- and cross-spectrum. This has surprising implications: when the $TT$ estimator is applied to a MV ILC of SO-LAT-like observations, the bias to the power spectrum of delensed $B$-modes is larger and has higher variance than in the case of no foreground cleaning, despite the bias to the template auto-spectrum being smaller and the bias to the cross-spectrum being practically unchanged --- this is a consequence of the cancellation between the two being spoilt.
    
    We use the Websky simulations to study in detail the case of SO --- ignoring polarized foregrounds --- and find that when delensing with a HO $TT$ estimator applied to CMB modes $l<3500$, the bias can be as large as $1.5\sigma$ before foreground cleaning, or $3\sigma$ when reconstructing from a MV ILC of temperature maps. Though it is in principle also possible for the $TE$ estimator to be biased by the foreground non-Gaussianity, we find this effect to be negligible for SO.

    In real analyses, the $TT$ estimator will likely be combined with other tracers --- other internal reconstructions, or perhaps external tracers. We have shown that when this is the case, the power spectrum of delensed $B$-modes receives new, non-trivial couplings beyond a simple dilution of the spurious power found in the case of a single tracer. Fortunately, these new terms appear to work in our favour, as they partially cancel with the diluted ones. For example, when the CIB is used in combination with $TT$ and other QEs to delens SO data, the cross-term cancels most of the diluted $\Delta C^{BB, \mathrm{res}}$, and allows the $l_{\mathrm{max}}$ in the $TT$ reconstruction to be larger than one would naively expect before receiving too large amounts of additional power and either incurring too large a bias, or degrading the delensing efficiency more than desired.

    The impact of foreground non-Gaussianity can be modeled very accurately using simple analytic expressions, as long as the lensing spectra these calculations rely on --- the auto-spectrum of $TT$ lensing reconstructions, and their cross-correlation with CMB lensing --- come from smooth fits to measurements that are themselves affected by the foreground non-Gaussianity. Once the non-Gaussian effects can be modeled, we are no longer obligated to prioritize bias mitigation over reconstruction efficiency (and thus lensing $B$-mode removal), and the goal becomes to determine what combination of analysis choices will result in the least amount of $B$-mode power after delensing, be it due to residual lensing or foreground-related contributions. 
    
    In this context, alternative $TT$ QEs such as the point-source-hardened or the shear-only estimator are likely to be crucial, as they are by design more robust to foregrounds than the Hu-Okamoto QE. On the one hand, they return reconstructions that lead to negligible delensing bias, even when the non-Gaussian effects are not modeled explicitly. On the other hand, when delensing with pipelines built around these estimators, the contribution from foreground non-Gaussianity to the power spectrum of delensed $B$-modes is much smaller than in the HO $TT$ case. As a result, the point-source-hardened estimator, which in general comes close to the HO $TT$ in terms of reconstruction efficiency, can ultimately yield the lowest amount of $B$-mode power after delensing. This is thanks to being able to extract information from smaller scales with little penalty in the form of additional foreground contributions, but also due to the fact that it can be applied to MV ILC maps without seeing its performance worsened.

    Given the nuanced interplay of effects that are taking place, with cancellations that can be highly beneficial, but which depend on the experimental configuration and can be easily spoilt by masking choices and foreground cleaning, it will be important to be able to estimate the biases ahead of any application on real data, such that the most optimal analysis choices can be made. This is particularly important whenever modeling the bias away is not an option --- in such case, the tradeoff between bias and noise of different mitigation methods should be thoroughly explored in a way that is specific to the experiment in question. Alternatively, when taking the modeling route, estimating the bias ahead of time would help determine the analysis strategy that minimizes the total $B$-mode power after delensing, as it is not clear a priori at what point the extra delensing efficiency obtained by a more audacious pipeline will be countered by the additional power sourced by foreground non-Gaussianity. This will entail validating pipelines on realistic simulations that accurately reproduce the foreground non-Gaussianity. Alternatively, it would be useful to have flexible analytic tools that can predict the biases as a function of the experimental and analysis parameters --- work on this front is in fact ongoing~\cite{ref:cosmoblender_in_prep}.
    
    Either approach will benefit from a better understanding of the extragalactic foregrounds. This is a critical stage of the path ahead: the uncertainty in the foreground modeling is likely to be the largest term in our error budget, but it is unclear how to account for it\footnote{This uncertainty is not expected to affect the accuracy of our empirically-calibrated prescription for modeling the bias to the delensed $B$-mode spectrum. It could only do so if the simulations grossly misrepresented the relative importance of the different couplings in appendix~\ref{appendix:TT_biases} --- this seems unlikely.}. The bias amplitudes we report are likely to be underestimated due to halo mass resolution of Websky, which misses contributions to the CIB from halos lighter than $10^{12}\,\mathrm{M}_{\odot}$ --- note, for instance, that the correlation between the 545\,GHz CIB map of Websky with CMB lensing, shown in figure~\ref{fig:multitracer_rho}, is a few percent lower than what was seen by~\cite{ref:sherwin_15}.
    
    In conclusion: we have shown that the challenge posed by non-Gaussian, extragalactic foregrounds can be overcome through modeling or bias mitigation strategies. Thanks to this,  the rich amounts of information encoded in the temperature anisotropies of the CMB can be leveraged to delens $B$-modes and improve constraints on the amplitude of primordial gravitational waves, a key open question in our quest to understand the Universe.

    \section*{Acknowledgements}
    We wish to thank Noah Sailer and Toshiya Namikawa for useful discussions regarding foreground removal; Emmanuel Schaan for advice on matched-filtering techniques; Marius Millea for useful comments on the masking process and for feedback on an earlier draft; Noah Weaverdyck for insights about LSS residuals in galaxy weights; Anthony Challinor for kindly providing comments on a draft and thoughts on the modeling approach; Alex van Engelen for feedback on a draft; and Blake Sherwin for discussions at an early stage on how to best isolate the delensing biases. We are also grateful to George Stein, Giuseppe Puglisi and Zack Li for assistance with the Websky simulations.
    
    In producing this work, we have made use of \texttt{numpy}~\cite{ref:numpy}, \texttt{scipy}~\cite{ref:scipy}, \texttt{matplotlib}~\cite{ref:matplotlib}, \texttt{pixell}\footnote{\url{https://github.com/simonsobs/pixell}}, \texttt{symlens}, \texttt{HEALPix}/\texttt{healpy}~\cite{Gorski:2004by, ref:healpy_paper} and \texttt{BasicILC}; we are grateful to their developers --- especially Mat Madhavacheril, Andrea Zonca and Emmanuel Schaan --- for making their code and tutorials available.
    
    ABL is a BCCP fellow at UC Berkeley and Lawrence Berkeley National Laboratory. SF is funded by the Physics Division of Lawrence Berkeley National Laboratory. 
    
    This work was carried out on the territory of xučyun (Huichin), the ancestral and unceded land of the Chochenyo speaking Ohlone people, the successors of the sovereign Verona Band of Alameda County.
    
    \onecolumngrid
    \appendix
    \section{$B$-mode delensing with a $TT$ QE --- possible biases} \label{appendix:TT_biases}
    Whenever lensing reconstructions are contaminated by non-Gaussian, extragalactic foregrounds, the angular spectra of $B$-modes delensed using those reconstructions will receive numerous new contributions. In equations~\eqref{eqn:bias_couplings_cross} and~\eqref{eqn:bias_couplings_auto}, we highlighted a subset of those term, the ones that we expect to be most significant based on arguments we will outline below. In this appendix, we provide the full list of contributions that are zero- or first-order in $\kappa$.
    
    For the sake of clarity, we take several notational shortcuts. In our presentation, the Gaussian contraction over fields denoted with the superscript `obs' and linked by over-bars is to be taken first, followed by the connected $n$-point function of the remaining $n$ fields inside the expectation value.
    
    The $n$-point functions can themselves be broken into different couplings. To explore these, we introduce the notation $\tilde{X}[X,\kappa]$, where $X$ is either $T$ or $E$, to represent the functional dependence of $\tilde{X}$ on the unlensed field $X$ and $\kappa$; recall that $\tilde{X}$ is linear in $X$, so where $\kappa$ is uncontracted, the unlensed field $X$ is implied. We then use another set of over-bars to denote which unlensed fields are coupled together inside the $n$-point function. On the other hand, and in order to highlight their importance in our investigation, we use bars under the expression to identify the foreground and convergence fields that are coupled together in bispectra and trispectra.

    We begin by considering the cross-correlation of lensing template and observed $B$-modes. At leading order, the only possible bias term is a function of the fully-connected $\langle\tilde{B}\tilde{E} \hat{\kappa}^{TT}\rangle$ trispectrum,
    \begin{equation}
    \bcontraction[1.5ex]{ \Delta C_l^{\tilde{B} \times  \hat{B}^{\mathrm{lens}}} \supset g_l \big[ \langle \tilde{B}[E,}{\kappa}{] \tilde{E}[E, {\kappa}] \hat{\kappa}^{TT}[}{s^{\mathrm{NG}}}
    \bcontraction[1.5ex]{ \Delta C_l^{\tilde{B} \times  \hat{B}^{\mathrm{lens}}} \supset g_l \big[ \langle \tilde{B}[E,\kappa] \tilde{E}[E, {\kappa}] \hat{\kappa}^{TT} [}{{s}^{\mathrm{NG}}}{, }{{s}^{\mathrm{NG}}}
    \wick[offset=1.5em]{  \Delta C_l^{\tilde{B} \times  \hat{B}^{\mathrm{lens}}} \supset g_l \left[ \langle \tilde{B}[\c1E,\kappa] \tilde{E}[\c1 {E}, \kappa] \hat{\kappa}^{TT} \left[ s^{\mathrm{NG}},  {s}^{\mathrm{NG}}\right] \rangle \right]}  \,.
    \end{equation}
    This is analogous to the bispectrum bias studied in the context of CMB lensing reconstructions (e.g.,~\cite{ref:van_engelen_et_al_14, ref:sailer_et_al_21}), and can be modeled using equation~\eqref{eqn:full_theory_cross}.
    
    When it comes to the auto-spectrum of the lensing $B$-modes template, many more bias couplings are in principle possible. First, there is the terms that feature a Gaussian contraction of the $E$-modes across templates. These are
    \begin{equation}\label{eqn:trispec_bias_appendix}
        \bcontraction[1.5ex]{  \Delta C_l^{\hat{B}^{\mathrm{lens}}\times  \hat{B}^{\mathrm{lens}}} \supset  h_l \big[ \langle  {E^{\mathrm{obs}}} \hat{\kappa}^{\mathrm{TT}} \big[  }{  s^{\mathrm{NG}}  }{  ,   }{  s^{\mathrm{NG}}  }
        \bcontraction[1.5ex]{  \Delta C_l^{\hat{B}^{\mathrm{lens}}\times  \hat{B}^{\mathrm{lens}}} \supset  h_l \big[ \langle  {E^{\mathrm{obs}}} \hat{\kappa}^{\mathrm{TT}} \big[ s^{\mathrm{NG}}, }{s^{\mathrm{NG}}}{  \big]  E^{\mathrm{obs}} \hat{\kappa}^{\mathrm{TT}} \big[  }{s^{\mathrm{NG}}}
        \bcontraction[1.5ex]{    \Delta C_l^{\hat{B}^{\mathrm{lens}}\times  \hat{B}^{\mathrm{lens}}} \supset  h_l \big[ \langle  {E^{\mathrm{obs}}} \hat{\kappa}^{\mathrm{TT}} \big[ s^{\mathrm{NG}}, s^{\mathrm{NG}} \big]  E^{\mathrm{obs}} \hat{\kappa}^{\mathrm{TT}} \big[  }{s^{\mathrm{NG}}}{,}{s^{\mathrm{NG}}}
        \Delta C_l^{\hat{B}^{\mathrm{lens}}\times  \hat{B}^{\mathrm{lens}}} \supset \wick[offset=1.5em]{ h_l \left[ \langle  \c {E^{\mathrm{obs}}} \hat{\kappa}^{\mathrm{TT}} \left[s^{\mathrm{NG}}, s^{\mathrm{NG}}\right]  \c {E^{\mathrm{obs}}} \hat{\kappa}^{\mathrm{TT}} \left[s^{\mathrm{NG}},s^{\mathrm{NG}}\right]\rangle \right]}\,,
    \end{equation}
    \begin{equation}\label{eqn:prim_bispec_bias_appendix}
        \bcontraction[1.5ex]{ \hphantom{h} \Delta C_l^{\hat{B}^{\mathrm{lens}}\times  \hat{B}^{\mathrm{lens}}} \supset  2\,h_l \big[ \langle  {E^{\mathrm{obs}}} \hat{\kappa}^{\mathrm{TT}} \big[ \tilde{T}[T,\kappa], \tilde{T}[T,}{   \kappa    }{  ] \big]  E^{\mathrm{obs}} \hat{\kappa}^{\mathrm{TT}} \big[  }{s^{\mathrm{NG}}}
        \bcontraction[1.5ex]{ \hphantom{h} \Delta C_l^{\hat{B}^{\mathrm{lens}}\times  \hat{B}^{\mathrm{lens}}} \supset 2\,h_l \big[ \langle  {E^{\mathrm{obs}}} \hat{\kappa}^{\mathrm{TT}} \big[ \tilde{T}[{T},\kappa], \tilde{T}[ T,\kappa]\big]  {E^{\mathrm{obs}}} \hat{\kappa}^{\mathrm{TT}} \big[   }{ s^{\mathrm{NG}} }{ , }{ s^{\mathrm{NG}} }
        \Delta C_l^{\hat{B}^{\mathrm{lens}}\times  \hat{B}^{\mathrm{lens}}} \supset \wick[offset=1.5em]{ 2\,h_l \left[ \langle  \c2 {E^{\mathrm{obs}}} \hat{\kappa}^{\mathrm{TT}} \left[ \tilde{T}[\c1{T},\kappa], \tilde{T}[\c1 T,\kappa]\right]  \c2 {E^{\mathrm{obs}}} \hat{\kappa}^{\mathrm{TT}} \left[s^{\mathrm{NG}},s^{\mathrm{NG}}\right]\rangle \right]}\,,
    \end{equation}
    and
    \begin{equation}\label{eqn:sec_bispec_bias_appendix}
        \bcontraction[1.5ex]{
        \Delta \,C_l^{\hat{B}^{\mathrm{lens}}\times  \hat{B}^{\mathrm{lens}}} \supset  4\,h_l \big[ \langle  {E^{\mathrm{obs}}} \hat{\kappa}^{\mathrm{TT}} \big[ \tilde{T}[T,
        }
        {
        \kappa
        }
        {
        ],
        }
        {
        s
        }
        \bcontraction[1.5ex]{
        \Delta \, C_l^{\hat{B}^{\mathrm{lens}}\times  \hat{B}^{\mathrm{lens}}} \supset  4\,h_l \big[ \langle  {E^{\mathrm{obs}}} \hat{\kappa}^{\mathrm{TT}} \big[ \tilde{T}[T,\kappa],
        }
        {
        s
        }
        {
        s^{\mathrm{NG}}\big]   {E^{\mathrm{obs}}} \hat{\kappa}^{\mathrm{TT}} \big[\tilde{T}[ {T},\kappa],
        }
        {
        s
        }
        \Delta C_l^{\hat{B}^{\mathrm{lens}}\times  \hat{B}^{\mathrm{lens}}} \supset \wick[offset=1.5em]{ 4\,h_l \left[ \langle  \c {E^{\mathrm{obs}}} \hat{\kappa}^{\mathrm{TT}} \big[ \tilde{T}[\c2 {T},\kappa], s^{\mathrm{NG}}\big]  \c {E^{\mathrm{obs}}} \hat{\kappa}^{\mathrm{TT}} \big[\tilde{T}[\c2 {T},\kappa],s^{\mathrm{NG}}\big]\rangle \right]} \,.
    \end{equation}
    Based on their coupling structure, we expect these to be the dominant bias terms. The reason is that this is the coupling arrangement in which the weights in the integrands are the least-tightly-coupled, leaving them free to explore and accumulate signal over the largest volume of multipole space (see the discussion in appendix~A of~\cite{ref:baleato_20_internal}). This view is supported by the fact that terms with this general structure are the only connected contributions to the template auto- that are included in the standard calculation of the residual lensing $B$-mode spectrum, equation~\eqref{eqn:clbbdel_theory}~\cite{ref:baleato_20_internal}.
    
    Terms~\eqref{eqn:trispec_bias_appendix}--\eqref{eqn:sec_bispec_bias_appendix} can be modeled using equation~\eqref{eqn:full_theory_auto}. The connection with the lensing reconstruction biases is then explicit: equation~\eqref{eqn:trispec_bias_appendix} is related to the trispectrum bias, equation~\eqref{eqn:prim_bispec_bias_appendix} to the primary bispectrum bias, and~\eqref{eqn:sec_bispec_bias_appendix} to the secondary bispectrum bias.
    
    There are also terms where the disconnected coupling is between the $E$-modes in one template and an observed $T$ field in the QE in that same template:
    \begin{equation}
        \bcontraction[1.5ex]{
        \Delta C_l^{\hat{B}^{\mathrm{lens}}\times  \hat{B}^{\mathrm{lens}}} \supset 8\,h_l \big[ \langle  {E^{\mathrm{obs}}} \hat{\kappa}^{\mathrm{TT}} \big[  {T}^{\mathrm{obs}},
        }{
        s^{\mathrm{NG}}
        }{
        \big]  \tilde{E}[ {E},
        }{
        \kappa
        }
        \bcontraction[1.5ex]{
        \Delta C_l^{\hat{B}^{\mathrm{lens}}\times  \hat{B}^{\mathrm{lens}}} \supset 8\,h_l \big[ \langle  {E^{\mathrm{obs}}} \hat{\kappa}^{\mathrm{TT}} \big[  {T}^{\mathrm{obs}}, s^{\mathrm{NG}} \big]  \tilde{E}[ {E},
        }{
        \kappa
        }{
        \hat{\kappa}^{\mathrm{TT}} \big[\tilde{T}[ {T},\kappa],
        }{
        s^{\mathrm{NG}}
        }
        \Delta C_l^{\hat{B}^{\mathrm{lens}}\times  \hat{B}^{\mathrm{lens}}} \supset \wick[offset=1.5em]{ 8\,h_l \left[ \langle  \c {E^{\mathrm{obs}}} \hat{\kappa}^{\mathrm{TT}} \big[ \c {T}^{\mathrm{obs}}, s^{\mathrm{NG}}\big]  \tilde{E}[ \c1 {E},\kappa] \hat{\kappa}^{\mathrm{TT}} \big[\tilde{T}[\c1 {T},\kappa],s^{\mathrm{NG}}\big]\rangle \right]}\,,
    \end{equation}
    \begin{equation}
        \bcontraction[1.5ex]{
        \Delta C_l^{\hat{B}^{\mathrm{lens}}\times  \hat{B}^{\mathrm{lens}}} \supset 8\,h_l \big[ \langle  {E^{\mathrm{obs}}} \hat{\kappa}^{\mathrm{TT}} \big[  {T}^{\mathrm{obs}}, 
        }{
        s^{\mathrm{NG}}
        }{
        \big]  \tilde{E}[  {E},\kappa] \hat{\kappa}^{\mathrm{TT}} \big[\tilde{T}[ {T},
        }{
        \kappa
        }
        \bcontraction[1.5ex]{
        \Delta C_l^{\hat{B}^{\mathrm{lens}}\times  \hat{B}^{\mathrm{lens}}} \supset 8\,h_l \big[ \langle  {E^{\mathrm{obs}}} \hat{\kappa}^{\mathrm{TT}} \big[  {T}^{\mathrm{obs}}, s^{\mathrm{NG}} \big]  \tilde{E}[  {E},\kappa] \hat{\kappa}^{\mathrm{TT}} \big[\tilde{T}[ {T},
        }{
        \kappa
        }{
        ],
        }{
        s^{\mathrm{NG}}
        }
        \Delta C_l^{\hat{B}^{\mathrm{lens}}\times  \hat{B}^{\mathrm{lens}}} \supset \wick[offset=1.5em]{ 8\,h_l \left[ \langle  \c {E^{\mathrm{obs}}} \hat{\kappa}^{\mathrm{TT}} \big[ \c {T}^{\mathrm{obs}}, s^{\mathrm{NG}}\big]  \tilde{E}[ \c1 {E},\kappa] \hat{\kappa}^{\mathrm{TT}} \big[\tilde{T}[\c1 {T},\kappa],s^{\mathrm{NG}}\big]\rangle \right]} \,,
    \end{equation}
    \begin{equation}
        \bcontraction[1.5ex]{
        \Delta C_l^{\hat{B}^{\mathrm{lens}}\times  \hat{B}^{\mathrm{lens}}} \supset 4\,h_l \big[ \langle  {E^{\mathrm{obs}}} \hat{\kappa}^{\mathrm{TT}} \big[ {T}^{\mathrm{obs}}, \tilde{T}[ {T},\kappa]\big]  \tilde{E}[ E,
        }{
        \kappa
        }{
        ] \hat{\kappa}^{\mathrm{TT}} \big[
        }{
        s^{\mathrm{NG}}
        }
        \bcontraction[1.5ex]{
        \Delta C_l^{\hat{B}^{\mathrm{lens}}\times  \hat{B}^{\mathrm{lens}}} \supset 4\,h_l \big[ \langle  {E^{\mathrm{obs}}} \hat{\kappa}^{\mathrm{TT}} \big[ {T}^{\mathrm{obs}}, \tilde{T}[ {T},\kappa]\big]  \tilde{E}[ E, \kappa ] \hat{\kappa}^{\mathrm{TT}} \big[
        }{
        s^{\mathrm{NG}}
        }{
        , 
        }{
        s^{\mathrm{NG}}
        }
        \Delta C_l^{\hat{B}^{\mathrm{lens}}\times  \hat{B}^{\mathrm{lens}}} \supset \wick[offset=1.5em]{ 4\,h_l \left[ \langle  \c {E^{\mathrm{obs}}} \hat{\kappa}^{\mathrm{TT}} \big[ \c {T}^{\mathrm{obs}}, \tilde{T}[\c1 {T},\kappa]\big]  \tilde{E}[ \c1 {E},\kappa] \hat{\kappa}^{\mathrm{TT}} \big[s^{\mathrm{NG}},s^{\mathrm{NG}}\big]\rangle \right]} \,,
    \end{equation}
    and
    \begin{equation}
        \bcontraction[1.5ex]{
        \Delta C_l^{\hat{B}^{\mathrm{lens}}\times  \hat{B}^{\mathrm{lens}}} \supset 4\,h_l \big[ \langle  {E^{\mathrm{obs}}} \hat{\kappa}^{\mathrm{TT}} \big[ {T}^{\mathrm{obs}}, \tilde{T}[ {T},
        }{
        \kappa
        }{
        ]\big]  \tilde{E}[ E, \kappa] \hat{\kappa}^{\mathrm{TT}} \big[
        }{
        s^{\mathrm{NG}}
        }
        \bcontraction[1.5ex]{
        \Delta C_l^{\hat{B}^{\mathrm{lens}}\times  \hat{B}^{\mathrm{lens}}} \supset 4\,h_l \big[ \langle  {E^{\mathrm{obs}}} \hat{\kappa}^{\mathrm{TT}} \big[ {T}^{\mathrm{obs}}, \tilde{T}[ {T},\kappa]\big]  \tilde{E}[ E, \kappa ] \hat{\kappa}^{\mathrm{TT}} \big[
        }{
        s^{\mathrm{NG}}
        }{
        , 
        }{
        s^{\mathrm{NG}}
        }
        \Delta C_l^{\hat{B}^{\mathrm{lens}}\times  \hat{B}^{\mathrm{lens}}} \supset \wick[offset=1.5em]{ 4\,h_l \left[ \langle  \c {E^{\mathrm{obs}}} \hat{\kappa}^{\mathrm{TT}} \big[ \c {T}^{\mathrm{obs}}, \tilde{T}[\c1 {T},\kappa]\big]  \tilde{E}[ \c1 {E},\kappa] \hat{\kappa}^{\mathrm{TT}} \big[s^{\mathrm{NG}},s^{\mathrm{NG}}\big]\rangle \right]} \,.
    \end{equation}
    Although these terms are more tightly coupled than those in~\eqref{eqn:trispec_bias_appendix} through~\eqref{eqn:sec_bispec_bias_appendix}, refs.~\cite{ref:baleato_20_internal} and~\cite{ref:namikawa_17} showed that couplings like these constitute the leading, pure-lensing corrections to equation~\eqref{eqn:clbbdel_theory}; they therefore have the potential to be somewhat relevant.

    Similarly, there are terms where the $E$-modes feature in a Gaussian contraction with the $T$ in the QE that appears in the \emph{other} leg of the correlation:
    \begin{equation}\label{eqn:tightliy_coupled_term_1}
        \bcontraction[1.5ex]{
        \Delta C_l^{\hat{B}^{\mathrm{lens}}\times  \hat{B}^{\mathrm{lens}}} \supset  8\,h_l \big[ \langle   {E^{\mathrm{obs}}}  \hat{\kappa}^{\mathrm{TT}} \big[\tilde{T}[ {T},\kappa],
        }{
        s^{\mathrm{NG}}
        }{
        \big] \tilde{E}[ {E},
        }{
        \kappa
        }
        \bcontraction[1.5ex]{
        \Delta C_l^{\hat{B}^{\mathrm{lens}}\times  \hat{B}^{\mathrm{lens}}} \supset  8\,h_l \big[ \langle   {E^{\mathrm{obs}}}  \hat{\kappa}^{\mathrm{TT}} \big[\tilde{T}[ {T},\kappa], s^{\mathrm{NG}}\big] \tilde{E}[ {E},
        }{
        \kappa
        }{
        ] \hat{\kappa}^{\mathrm{TT}} \big[ \c2 {T}^{\mathrm{obs}},
        }{
        s
        }
        \Delta C_l^{\hat{B}^{\mathrm{lens}}\times  \hat{B}^{\mathrm{lens}}} \supset \wick[offset=1.5em]{ 8\,h_l \left[ \langle  \c2 {E^{\mathrm{obs}}}  \hat{\kappa}^{\mathrm{TT}} \big[\tilde{T}[\c1 {T},\kappa],s^{\mathrm{NG}}\big] \tilde{E}[ \c1 {E},\kappa] \hat{\kappa}^{\mathrm{TT}} \big[ \c2 {T}^{\mathrm{obs}}, s^{\mathrm{NG}}\big]\rangle \right]} \,,
    \end{equation}
    \begin{equation}
        \bcontraction[1.5ex]{
        \Delta C_l^{\hat{B}^{\mathrm{lens}}\times  \hat{B}^{\mathrm{lens}}} \supset  8\,h_l \big[ \langle   {E^{\mathrm{obs}}}  \hat{\kappa}^{\mathrm{TT}} \big[\tilde{T}[ {T},
        }{
        \kappa 
        }{
        ],
        }{
        s^{\mathrm{NG}}
        }
        \bcontraction[1.5ex]{
        \Delta C_l^{\hat{B}^{\mathrm{lens}}\times  \hat{B}^{\mathrm{lens}}} \supset  8\,h_l \big[ \langle   {E^{\mathrm{obs}}}  \hat{\kappa}^{\mathrm{TT}} \big[\tilde{T}[ {T},\kappa], 
        }{
        s^{\mathrm{NG}}
        }{
        \big] \tilde{E}[ {E}, \kappa] \hat{\kappa}^{\mathrm{TT}} \big[ \c2 {T}^{\mathrm{obs}},
        }{
        s
        }
        \Delta C_l^{\hat{B}^{\mathrm{lens}}\times  \hat{B}^{\mathrm{lens}}} \supset \wick[offset=1.5em]{ 8\,h_l \left[ \langle  \c2 {E^{\mathrm{obs}}}  \hat{\kappa}^{\mathrm{TT}} \big[\tilde{T}[\c1 {T},\kappa],s^{\mathrm{NG}}\big] \tilde{E}[ \c1 {E},\kappa] \hat{\kappa}^{\mathrm{TT}} \big[ \c2 {T}^{\mathrm{obs}}, s^{\mathrm{NG}}\big]\rangle \right]} \,,
    \end{equation}
    \begin{equation}
        \bcontraction[1.5ex]{
        \Delta C_l^{\hat{B}^{\mathrm{lens}}\times  \hat{B}^{\mathrm{lens}}} \supset 4\,h_l \big[ \langle  {E^{\mathrm{obs}}} \hat{\kappa}^{\mathrm{TT}} \big[
        }{
        s^{\mathrm{NG}}
        }{
        , 
        }{
        s^{\mathrm{NG}}
        }
        \bcontraction[1.5ex]{
        \Delta C_l^{\hat{B}^{\mathrm{lens}}\times  \hat{B}^{\mathrm{lens}}} \supset 4\,h_l \big[ \langle  {E^{\mathrm{obs}}} \hat{\kappa}^{\mathrm{TT}} \big[s^{\mathrm{NG}},
        }{
        s^{\mathrm{NG}}
        }{
        \big] \tilde{E}[  {E},
        }{
        \kappa
        }
        \Delta C_l^{\hat{B}^{\mathrm{lens}}\times  \hat{B}^{\mathrm{lens}}} \supset \wick[offset=1.5em]{ 4\,h_l \left[ \langle  \c {E^{\mathrm{obs}}} \hat{\kappa}^{\mathrm{TT}} \big[s^{\mathrm{NG}},s^{\mathrm{NG}}\big] \tilde{E}[ \c2 {E},\kappa]  \hat{\kappa}^{\mathrm{TT}} \big[ \c {T}^{\mathrm{obs}}, \tilde{T}[\c2 {T},\kappa]\big] \rangle \right]} \,
    \end{equation}
    and
    \begin{equation}
        \bcontraction[1.5ex]{
        \Delta C_l^{\hat{B}^{\mathrm{lens}}\times  \hat{B}^{\mathrm{lens}}} \supset 4\,h_l \big[ \langle  {E^{\mathrm{obs}}} \hat{\kappa}^{\mathrm{TT}} \big[
        }{
        s^{\mathrm{NG}}
        }{
        , 
        }{
        s^{\mathrm{NG}}
        }
        \bcontraction[1.5ex]{
        \Delta C_l^{\hat{B}^{\mathrm{lens}}\times  \hat{B}^{\mathrm{lens}}} \supset 4\,h_l \big[ \langle  {E^{\mathrm{obs}}} \hat{\kappa}^{\mathrm{TT}} \big[s^{\mathrm{NG}},
        }{
        s^{\mathrm{NG}}
        }{
        \big] \tilde{E}[  {E},\kappa]  \hat{\kappa}^{\mathrm{TT}} \big[  {T}^{\mathrm{obs}}, \tilde{T}[ {T},
        }{
        \kappa
        }
        \Delta C_l^{\hat{B}^{\mathrm{lens}}\times  \hat{B}^{\mathrm{lens}}} \supset \wick[offset=1.5em]{ 4\,h_l \left[ \langle  \c {E^{\mathrm{obs}}} \hat{\kappa}^{\mathrm{TT}} \big[s^{\mathrm{NG}},s^{\mathrm{NG}}\big] \tilde{E}[ \c2 {E},\kappa]  \hat{\kappa}^{\mathrm{TT}} \big[ \c {T}^{\mathrm{obs}}, \tilde{T}[\c2 {T},\kappa]\big] \rangle \right]} \,.
    \end{equation}

    Finally, the Gaussian contraction can be among $T$ fields, either inside the same QE, as in
    \begin{equation}
        \bcontraction[1.5ex]{
        \Delta C_l^{\hat{B}^{\mathrm{lens}}\times  \hat{B}^{\mathrm{lens}}} \supset 2\,h_l \big[ \langle  \tilde{E} [ {E},\kappa]\hat{\kappa}^{\mathrm{TT}} \big[ {T}^{\mathrm{obs}},  {T}^{\mathrm{obs}}\big] \tilde{E}[ {E},
        }{
        \kappa
        }{
        ]  \hat{\kappa}^{\mathrm{TT}} \big[
        }{
        s^{\mathrm{NG}}
        }
        \bcontraction[1.5ex]{
        \Delta C_l^{\hat{B}^{\mathrm{lens}}\times  \hat{B}^{\mathrm{lens}}} \supset 2\,h_l \big[ \langle  \tilde{E} [ {E},\kappa]\hat{\kappa}^{\mathrm{TT}} \big[ {T}^{\mathrm{obs}},  {T}^{\mathrm{obs}}\big] \tilde{E}[ {E}, \kappa ]  \hat{\kappa}^{\mathrm{TT}} \big[
        }{
        s^{\mathrm{NG}}
        }{
        ,
        }{
        s^{\mathrm{NG}}
        }
        \Delta C_l^{\hat{B}^{\mathrm{lens}}\times  \hat{B}^{\mathrm{lens}}} \supset \wick[offset=1.5em]{2\,h_l \left[ \langle  \tilde{E} [\c2 {E},\kappa]\hat{\kappa}^{\mathrm{TT}} \big[\c {T}^{\mathrm{obs}}, \c {T}^{\mathrm{obs}}\big] \tilde{E}[ \c2 {E},\kappa]  \hat{\kappa}^{\mathrm{TT}} \big[ s^{\mathrm{NG}},s^{\mathrm{NG}}]\big] \rangle \right]} \,,
    \end{equation}
    and
    \begin{equation}
        \bcontraction[1.5ex]{
        \Delta C_l^{\hat{B}^{\mathrm{lens}}\times  \hat{B}^{\mathrm{lens}}} \supset 2\,h_l \big[ \langle  \tilde{E} [ {E},
        }{
        \kappa
        }{
        ]\hat{\kappa}^{\mathrm{TT}} \big[ {T}^{\mathrm{obs}},  {T}^{\mathrm{obs}}\big] \tilde{E}[ {E}, \kappa]  \hat{\kappa}^{\mathrm{TT}} \big[
        }{
        s^{\mathrm{NG}}
        }
        \bcontraction[1.5ex]{
        \Delta C_l^{\hat{B}^{\mathrm{lens}}\times  \hat{B}^{\mathrm{lens}}} \supset 2\,h_l \big[ \langle  \tilde{E} [ {E},\kappa]\hat{\kappa}^{\mathrm{TT}} \big[ {T}^{\mathrm{obs}},  {T}^{\mathrm{obs}}\big] \tilde{E}[ {E}, \kappa ]  \hat{\kappa}^{\mathrm{TT}} \big[
        }{
        s^{\mathrm{NG}}
        }{
        ,
        }{
        s^{\mathrm{NG}}
        }
        \Delta C_l^{\hat{B}^{\mathrm{lens}}\times  \hat{B}^{\mathrm{lens}}} \supset \wick[offset=1.5em]{2\,h_l \left[ \langle  \tilde{E} [\c2 {E},\kappa]\hat{\kappa}^{\mathrm{TT}} \big[\c {T}^{\mathrm{obs}}, \c {T}^{\mathrm{obs}}\big] \tilde{E}[ \c2 {E},\kappa]  \hat{\kappa}^{\mathrm{TT}} \big[ s^{\mathrm{NG}},s^{\mathrm{NG}}]\big] \rangle \right]} \,;
    \end{equation}
    or between QEs in different legs of the correlator,
    \begin{equation}\label{eqn:tightliy_coupled_term_last}
        \bcontraction[1.5ex]{
        \Delta C_l^{\hat{B}^{\mathrm{lens}}\times  \hat{B}^{\mathrm{lens}}} \supset 8\,h_l \big[ \langle  \tilde{E} [{E},\kappa]\hat{\kappa}^{\mathrm{TT}} \big[ {T}^{\mathrm{obs}},
        }{
        s^{\mathrm{NG}}
        }{
        \big] \tilde{E}[ {E},
        }{
        \kappa
        }
        \bcontraction[1.5ex]{
        \Delta C_l^{\hat{B}^{\mathrm{lens}}\times  \hat{B}^{\mathrm{lens}}} \supset 8\,h_l \big[ \langle  \tilde{E} [{E},\kappa]\hat{\kappa}^{\mathrm{TT}} \big[ {T}^{\mathrm{obs}}, s^{\mathrm{NG}}\big] \tilde{E}[ {E},
        }{
        \kappa
        }{
        ]  \hat{\kappa}^{\mathrm{TT}} \big[ {T}^{\mathrm{obs}},
        }{
        s^{\mathrm{NG}}
        }
        \Delta C_l^{\hat{B}^{\mathrm{lens}}\times  \hat{B}^{\mathrm{lens}}} \supset \wick[offset=1.5em]{8\,h_l \left[ \langle  \tilde{E} [\c2{E},\kappa]\hat{\kappa}^{\mathrm{TT}} \big[\c {T}^{\mathrm{obs}},  s^{\mathrm{NG}}\big] \tilde{E}[ \c2 {E},\kappa]  \hat{\kappa}^{\mathrm{TT}} \big[ \c {T}^{\mathrm{obs}},s^{\mathrm{NG}}]\big] \rangle \right]} \,.
    \end{equation}
    Terms~\eqref{eqn:tightliy_coupled_term_1}--\eqref{eqn:tightliy_coupled_term_last} are all very tightly-coupled, so we expect them to make only small contributions to the total bias.

    \section{$B$-mode delensing with a $TE$ QE --- possible biases} \label{appendix:TE_biases}
    We can use the notation and arguments described in appendix~\ref{appendix:TT_biases} to dissect the possible biases that appear when delensing with a $TE$ estimator. The least tightly-coupled --- and thus most concerning --- term is
    \begin{equation}
        \bcontraction[1.5ex]{
        \Delta C_l^{\hat{B}^{\mathrm{lens}}\times  \hat{B}^{\mathrm{lens}}} \supset  2\,h_l \big[ \langle   {E^{\mathrm{obs}}} \hat{\kappa}^{\mathrm{TE}} \big[
        }{
        s^{\mathrm{NG}}
        }{
        , \tilde{E}[{E},
        }{
        \kappa
        }
        \bcontraction[1.5ex]{
        \Delta C_l^{\hat{B}^{\mathrm{lens}}\times  \hat{B}^{\mathrm{lens}}} \supset  2\,h_l \big[ \langle   {E^{\mathrm{obs}}} \hat{\kappa}^{\mathrm{TE}} \big[s^{\mathrm{NG}}, \tilde{E}[{E},
        }{
        \kappa
        }{
        ]\big]  {E^{\mathrm{obs}}} \hat{\kappa}^{\mathrm{TE}} \big[
        }{
        s^{\mathrm{NG}}
        }
        \Delta C_l^{\hat{B}^{\mathrm{lens}}\times  \hat{B}^{\mathrm{lens}}} \supset \wick[offset=1.5em]{ 2\,h_l \left[ \langle  \c {E^{\mathrm{obs}}} \hat{\kappa}^{\mathrm{TE}} \big[ s^{\mathrm{NG}}, \tilde{E}[\c2 {E},\kappa]\big]  \c {E^{\mathrm{obs}}} \hat{\kappa}^{\mathrm{TE}} \big[s^{\mathrm{NG}}, \tilde{E}[\c2 {E},\kappa]\big]\rangle \right]} \,.
    \end{equation}
    This is essentially a `secondary bispectrum' bias that can be modeled via equation~\eqref{eqn:auto_theory} once the bias to the $\hat{\kappa}^{TE}$ auto-spectrum is known. Other, likely smaller contributions are
    \begin{equation}
        \bcontraction[1.5ex]{
        \Delta C_l^{\hat{B}^{\mathrm{lens}}\times  \hat{B}^{\mathrm{lens}}} \supset  2\,h_l \big[ \langle   {E^{\mathrm{obs}}} \hat{\kappa}^{\mathrm{TE}} \big[
        }{
        s^{\mathrm{NG}}
        }{
        ,  {E}^{\mathrm{obs}}\big]  \tilde{E}[ {E},\kappa] \hat{\kappa}^{\mathrm{TE}} \big[
        }{
        s^{\mathrm{NG}}
        }
        \bcontraction[1.5ex]{
        \Delta C_l^{\hat{B}^{\mathrm{lens}}\times  \hat{B}^{\mathrm{lens}}} \supset  2\,h_l \big[ \langle   {E^{\mathrm{obs}}} \hat{\kappa}^{\mathrm{TE}} \big[ s^{\mathrm{NG}},  {E}^{\mathrm{obs}}\big]  \tilde{E}[ {E},\kappa] \hat{\kappa}^{\mathrm{TE}} \big[
        }{
        s^{\mathrm{NG}}
        }{
        , \tilde{E}[ {E},
        }{
        \kappa
        }
        \Delta C_l^{\hat{B}^{\mathrm{lens}}\times  \hat{B}^{\mathrm{lens}}} \supset \wick[offset=1.5em]{ 2\,h_l \left[ \langle  \c {E^{\mathrm{obs}}} \hat{\kappa}^{\mathrm{TE}} \big[ s^{\mathrm{NG}}, \c {E}^{\mathrm{obs}}\big]  \tilde{E}[ \c1 {E},\kappa] \hat{\kappa}^{\mathrm{TE}} \big[s^{\mathrm{NG}}, \tilde{E}[\c1 {E},\kappa]\big]\rangle \right]} \,,
    \end{equation}
    and
    \begin{equation}
        \bcontraction[1.5ex]{
        \Delta C_l^{\hat{B}^{\mathrm{lens}}\times  \hat{B}^{\mathrm{lens}}} \supset  2\,h_l \big[ \langle   {E^{\mathrm{obs}}} \hat{\kappa}^{\mathrm{TE}} \big[
        }{
        s^{\mathrm{NG}}
        }{
        ,  {E}^{\mathrm{obs}}\big]  \tilde{E}[ {E},
        }{
        \kappa
        }
        \bcontraction[1.5ex]{
        \Delta C_l^{\hat{B}^{\mathrm{lens}}\times  \hat{B}^{\mathrm{lens}}} \supset  2\,h_l \big[ \langle   {E^{\mathrm{obs}}} \hat{\kappa}^{\mathrm{TE}} \big[ s^{\mathrm{NG}},  {E}^{\mathrm{obs}}\big]  \tilde{E}[ {E},
        }{
        \kappa
        }{
        ] \hat{\kappa}^{\mathrm{TE}} \big[
        }{
        s^{\mathrm{NG}}
        }
        \Delta C_l^{\hat{B}^{\mathrm{lens}}\times  \hat{B}^{\mathrm{lens}}} \supset \wick[offset=1.5em]{ 2\,h_l \left[ \langle  \c {E^{\mathrm{obs}}} \hat{\kappa}^{\mathrm{TE}} \big[ s^{\mathrm{NG}}, \c {E}^{\mathrm{obs}}\big]  \tilde{E}[ \c1 {E},\kappa] \hat{\kappa}^{\mathrm{TE}} \big[s^{\mathrm{NG}}, \tilde{E}[\c1 {E},\kappa]\big]\rangle \right]} \,,
    \end{equation}
    in which the Gaussian contraction is between $E$-modes that feature explictly in the template and $E$-modes that are input into the $TE$ QE in that same template. In addition to these, there is also
    \begin{equation}
        \bcontraction[1.5ex]{
        \Delta C_l^{\hat{B}^{\mathrm{lens}}\times  \hat{B}^{\mathrm{lens}}} \supset  4\,h_l \big[ \langle  \tilde{E}[ {E},\kappa] \hat{\kappa}^{\mathrm{TE}} \big[ 
        }{
        s^{\mathrm{NG}}
        }{
        ,  {E}^{\mathrm{obs}}\big]  \tilde{E}[  {E},
        }{
        \kappa
        }
        \bcontraction[1.5ex]{
        \Delta C_l^{\hat{B}^{\mathrm{lens}}\times  \hat{B}^{\mathrm{lens}}} \supset  4\,h_l \big[ \langle  \tilde{E}[ {E},\kappa] \hat{\kappa}^{\mathrm{TE}} \big[ s^{\mathrm{NG}},  {E}^{\mathrm{obs}}\big]  \tilde{E}[  {E},
        }{
        \kappa
        }{
        ] \hat{\kappa}^{\mathrm{TE}} \big[
        }{
        s^{\mathrm{NG}}
        }
        \Delta C_l^{\hat{B}^{\mathrm{lens}}\times  \hat{B}^{\mathrm{lens}}} \supset \wick[offset=1.5em]{ 4\,h_l \left[ \langle  \tilde{E}[\c {E},\kappa] \hat{\kappa}^{\mathrm{TE}} \big[ s^{\mathrm{NG}}, \c2 {E}^{\mathrm{obs}}\big]  \tilde{E}[ \c {E},\kappa] \hat{\kappa}^{\mathrm{TE}} \big[s^{\mathrm{NG}}, \c2 {E}^{\mathrm{obs}}\big]\rangle \right]} \,,
    \end{equation}
    where the Gaussian contraction is between $E$-modes across the two $TE$ QEs.
    
    \bibliography{main}
    
\end{document}